\documentclass[aps,10pt,prb,twocolumn, superscriptaddress]{revtex4}
\usepackage{graphicx}
\usepackage{amssymb, amsmath}
\usepackage{amsfonts}
\usepackage{color}
\usepackage{epstopdf}
\usepackage{multirow}
\usepackage{bm}
\newcommand\varpm{\mathbin{\vcenter{\hbox{%
  \oalign{\hfil$\scriptstyle+$\hfil\cr
          \noalign{\kern-.3ex}
          $\scriptscriptstyle({-})$\cr}%
}}}}
\newcommand\varmp{\mathbin{\vcenter{\hbox{%
  \oalign{$\scriptstyle({+})$\cr
          \noalign{\kern-.3ex}
          \hfil$\scriptscriptstyle-$\hfil\cr}%
}}}}
\DeclareMathAlphabet      {\mathbf}{OT1}{cmr}{bx}{n}

\begin{document}

\title{Magnetic structure and high-field magnetization of the distorted kagome lattice antiferromagnet Cs$_2$Cu$_3$SnF$_{12}$}
\author{K.~Matan}
\email[Corresponding author: ]{kittiwit.mat@mahidol.ac.th}
\affiliation{Department~of~Physics,~Faculty~of~Science,~Mahidol~University,~272~Rama~VI~Rd.,~Ratchathewi, Bangkok 10400, Thailand}
\affiliation{ThEP,~Commission of Higher Education,~328 Si Ayuthaya Road,~Bangkok 10400, Thailand}
\author{T.~Ono}
\affiliation{Department of Physical Science, School of Science, Osaka Prefecture University, Sakai, Osaka 599-8531, Japan}
\author{G.~Gitgeatpong}
\affiliation{Department~of~Physics,~Faculty~of~Science,~Mahidol~University,~272~Rama~VI~Rd.,~Ratchathewi, Bangkok 10400, Thailand}
\affiliation{ThEP,~Commission of Higher Education,~328 Si Ayuthaya Road,~Bangkok 10400, Thailand}
\affiliation{Department of Physics, Faculty of Science and Technology, Phranakhon Rajabhat University, Bangkok 10220, Thailand}
\author{K.~de~Roos}
\altaffiliation{Current address: Radboud University, Institute for Molecules and Materials, Heyendaalseweg 135, 6525 AJ Nijmegen, The Netherlands.}
\affiliation{Department~of~Physics,~Faculty~of~Science,~Mahidol~University,~272~Rama~VI~Rd.,~Ratchathewi, Bangkok 10400, Thailand}
\author{P. Miao}
\affiliation{Neutron Science Laboratory, Institute of Materials Structure Science, High Energy Accelerator Research Organization (KEK), 1-1 Oho, Tsukuba, Ibaraki 305-0801, Japan}
\author{S.~Torii}
\affiliation{Neutron Science Laboratory, Institute of Materials Structure Science, High Energy Accelerator Research Organization (KEK), 1-1 Oho, Tsukuba, Ibaraki 305-0801, Japan}
\author{T.~Kamiyama}
\affiliation{Neutron Science Laboratory, Institute of Materials Structure Science, High Energy Accelerator Research Organization (KEK), 1-1 Oho, Tsukuba, Ibaraki 305-0801, Japan}
\author{A.~Miyata}
\affiliation{Institute for Solid State Physics, The University of Tokyo, Kashiwa, Chiba 277-8581, Japan}
\author{A.~Matsuo}
\affiliation{Institute for Solid State Physics, The University of Tokyo, Kashiwa, Chiba 277-8581, Japan}
\author{K.~Kindo}
\affiliation{Institute for Solid State Physics, The University of Tokyo, Kashiwa, Chiba 277-8581, Japan}
\author{S.~Takeyama}
\affiliation{Institute for Solid State Physics, The University of Tokyo, Kashiwa, Chiba 277-8581, Japan}
\author{Y.~Nambu} 
\affiliation{Institute for Materials Research, Tohoku University, Sendai 980-8577, Japan}
\author{P.~Piyawongwatthana}
\affiliation{Institute of Multidisciplinary Research for Advanced Materials, Tohoku University, 2-1-1 Katahira, Sendai, Miyagi 980-8577, Japan}
\author{T.~J.~Sato}
\affiliation{Institute of Multidisciplinary Research for Advanced Materials, Tohoku University, 2-1-1 Katahira, Sendai, Miyagi 980-8577, Japan}
\author{H.~Tanaka}
\affiliation{Department of Physics, Tokyo Institute of Technology, Meguro-ku, Tokyo 152-8551, Japan}

\date{\today}
\begin{abstract}
High-resolution time-of-flight powder neutron diffraction and high-field magnetization were measured to investigate the magnetic structure and existence of a field-induced magnetic phase transition in the distorted kagome antiferromagnet Cs$_2$Cu$_3$SnF$_{12}$.  Upon cooling from room temperature, the compound undergoes a structural phase transition at $T_\textrm{t}=185$~K from the rhombohedral space group $R\bar{3}m$ with the perfect kagome spin network to the monoclinic space group $P2_1/n$ with the distorted kagome planes.   The distortion results in three inequivalent exchange interactions among the $S=1/2$ Cu$^{2+}$ spins that magnetically order below $T_\textrm{N}=20.2$~K.  Magnetization measured with a magnetic field applied within the kagome plane reveals small in-plane ferromagnetism resulting from spin canting.  On the other hand, the out-of-plane magnetization does not show a clear hysteresis loop of the ferromagnetic component nor a prominent anomaly up to 170 T, with the exception of the subtle knee-like bend around 90 T, which could indicate the 1/3 magnetization plateau. The combined analysis using the irreducible representations of the magnetic space groups and magnetic structure refinement on the neutron powder diffraction data suggests that the magnetic moments order in the magnetic space group $P2_1'/n'$ with the all-in-all-out spin structure, which by symmetry allows for the in-plane canting, consistent with the in-plane ferromagnetism observed in the magnetization. 
\end{abstract}
\maketitle

\begin{center}\rule{0.9\columnwidth}{0.4pt}\end{center}
\noindent\emph{Note added (2026).}\ \ The magnetic structure reported in this paper is incorrect.  In the {\scshape FullProf} refinements the ordered moments of the two crystallographically inequivalent Cu$^{2+}$ sites were placed in two \emph{separate} magnetic phases, so that their magnetic Bragg intensities were added incoherently; both sites belong to a single $\mathbf{k}=0$ magnetic structure, so their magnetic structure factors must be added \emph{coherently}.  The corrected refinements favor the coplanar 120$^\circ$ structure of magnetic space group $P2_1/n$ with negative vector chirality and ordered moment 0.52(2)~$\mu_\textrm{B}$, and the identification of the all-in-all-out structure of $P2_1'/n'$ is withdrawn.  All other results of this paper are unaffected.  The full corrected analysis is given in the Erratum reproduced at the end of this document, published as Phys.~Rev.~B \textbf{114}, 139902(E) (2026).
\begin{center}\rule{0.9\columnwidth}{0.4pt}\end{center}

\section{Introduction} 
A search for a quantum spin liquid state,  a strongly correlated state with a macroscopic degree of entanglement~\cite{Balents:2010ds, Savary:2017fk}, has been at the forefront of research in condensed matter physics.  One prime candidate with the possibility of hosting the quantum spin liquid is the kagome lattice antiferromagnet.  The unique mosaic of the kagome lattice decorated by corner sharing triangles gives rise to geometric frustration when the exchange interaction between the adjacent spins is antiferromagnetic.  This frustration, which is caused by the incompatibility between the underlying global geometry of the lattice and the local spin-spin interaction, gives rise to a macroscopic degenerate ground state, resulting in the suppression of a N\'eel ordered state and in the possible emergence of the quantum spin liquid state. The current consensus based on theoretical work~\cite{Yan1173,PhysRevLett.109.067201,Jiang:2012dw,PhysRevB.95.235107,PhysRevLett.118.137202,Zhueaat5535} suggests the quantum spin liquid as the ground state of the kagome lattice antiferromagnet.  However, the nature of this quantum spin liquid is not yet conclusive. Experimentally, some of the most studied systems include herbertsmithite~\cite{Shores:2005de, PhysRevLett.98.107204, PhysRevLett.104.147201, Han:2012fo} as well as its relatives~\cite{RevModPhys.88.041002}, volborthite~\cite{Hiroi:2001vy, doi:10.1143/JPSJ.78.033701}, vesignieite~\cite{Hiroi2009}, and jarosite~\cite{PhysRevB.61.6156, Inami.PhysRevB.61.12181, Grohol:2005bg, Matan:2006fl}, to name a few. However, all of these realizations are not ideal as they are plagued by further-neighbor interactions, spatially nonuniform exchange interactions due to lattice distortion, anisotropic interaction, and magnetic/non-magnetic inter-site defects.  As a result, experimentalists have, so far, been able to study these approximate realizations, the majority of which have the N\'eel ground state, and look for hints of hidden physics of the ideal system.  

\begin{figure*}
\centering \vspace{0in}
\includegraphics[width=17cm]{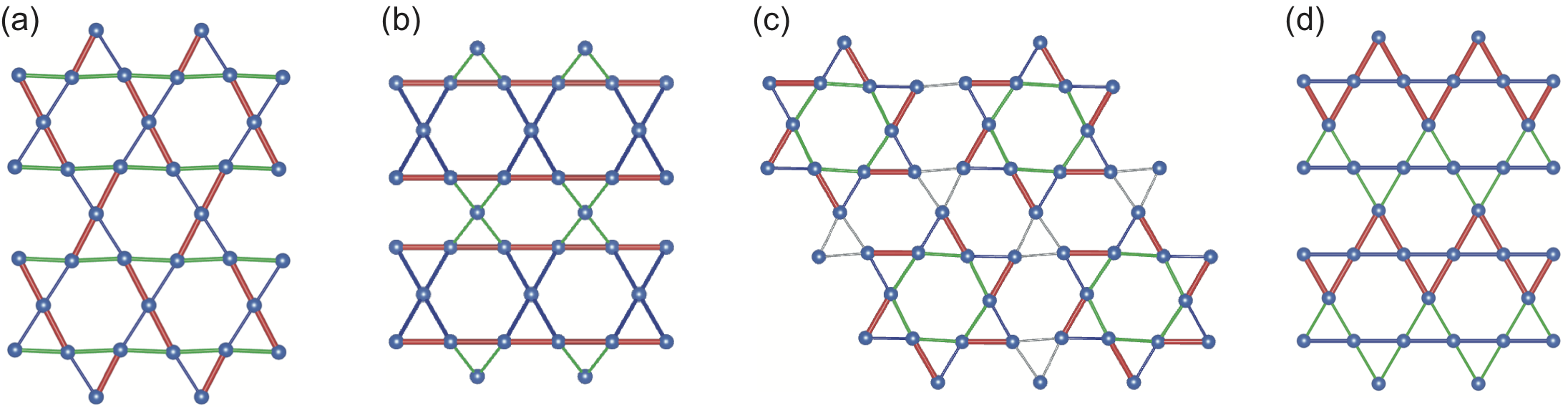}
\caption{(Color online) The distorted kagome planes are shown for (a) the low-temperature phase of Cs$_2$Cu$_3$SnF$_{12}$ ($P2_1/n$), (b) the low-temperature phase of Cs$_2$Cu$_3$CeF$_{12}$ ($Pnnn$), (c) the room-temperature phase of Rb$_2$Cu$_3$SnF$_{12}$ ($R\bar{3}$), and (d) the low-temperature phase of Cs$_2$Cu$_3$ZrF$_{12}$ ($P2_1/m$).  The different color bonds represent different nearest-neighbor exchange interactions resulting from the lattice distortion.}\label{fig1}
\end{figure*}

One of the approximate systems is the family of compounds $A_2$Cu$_3B$F$_{12}$, where $A$ is a monovalent ion Cs or Rb, and $B$ is a tetravalent ion Sn, Ce, Hf, Zr, or Ti.  At room temperature, some of the compounds in this family are reported to crystalize in the rhombohedral space group $R\bar{3}m$, in which the Cu$^{2+}$ ions form a perfect kagome plane comprising equilateral triangles, whereas in the other compounds they form a distorted kagome lattice.  For example, the room-temperature crystal structure of Cs$_2$Cu$_3$CeF$_{12}$ is orthorhombic $Pnnn$, where the Cu$^{2+}$ ions form a buckled kagome lattice, which can be viewed as two subsystems of alternating spin chains and dangling spins as shown in Fig.~\ref{fig1}(b)~\cite{PhysRevB.80.100406}.  The magnetic $S=1/2$ Cu$^{2+}$ spins order below $T_\textrm{N}=3$~K.  Even though the magnetic structure at zero magnetic field is unresolved, the magnetization measured up to 7 T with the applied field parallel to the crystallographic $a$-axis displays the 1/3 magnetization plateau suggesting the ferromagnetic alignment of the dangling spins (one third of the total number of spins) and presumably the antiferromagnetic alignment for spins along the chains (two thirds). 

Rb$_2$Cu$_3$SnF$_{12}$ crystalizes to the rhombohedral space group $R\bar{3}$ at room temperature with the $2a\times2a$ enlarged in-plane unit cell, as shown in Fig.~\ref{fig1}(c), compared to the undistorted kagome lattice. The lattice distortion at room temperature gives rise to four distinct exchange parameters, $J_1>J_2>J_3>J_4$. The system undergoes a further structural phase transition to the triclinic structure, space group $P\bar{1}$ at 215 K~\cite{Downie:2014jg}. The largest exchange interaction $J_1$ forms the pinwheel-like spin network.  A pair of spins connected by $J_1$ forms a singlet (dimer) of $S=0$, giving rise to the non-magnetic quantum ground state.  These dimers decorate the lattice forming the so-called pinwheel valence-bond-solid state~\cite{Morita}.  Inelastic neutron scattering has confirmed the single-to-triplet excitations~\cite{Matan:2010ha} and $^{63,65}$Cu nuclear magnetic resonance has revealed the field-dependence of the spin gap up to 30~T~\cite{PhysRevLett.110.247203}.  Furthermore, all relevant microscopic spin Hamiltonian parameters were extracted from the dispersion of the single-to-triplet excitations~\cite{Matan:2010ha, Matan:2014cn}.  The low-temperature-phase crystal structure of Rb$_2$Cu$_3$SnF$_{12}$ remains unresolved, for which the studies on a powder and single crystal sample give contradictory results~\cite{Downie:2013cz}.  For Rb$_2$Cu$_3$TiF$_{12}$, a recent study shows that at room temperature it crystallizes in the triclinic space group $P\bar{1}$ with a possible structural transition at higher temperature to a high symmetry phase probably of the space group $R\bar{3}$, as with Rb$_2$Cu$_3$SnF$_{12}$.  However, in contrast to Rb$_2$Cu$_3$SnF$_{12}$, Rb$_2$Cu$_3$TiF$_{12}$ displays a magnetically ordered state at low temperature~\cite{Downie:2015bj}. 

At room temperature, Cs$_2$Cu$_3$HfF$_{12}$ and Cs$_2$Cu$_3$ZrF$_{12}$ crystalize in the rhombohedral space group $R\bar{3}m$, where the Cu$^{2+}$ ions form a perfect kagome plane made up of equilateral triangles~\cite{Muller:1995im}.  However, at low temperature, they undergo a structural phase transition to a monoclinic space group with lower symmetry.  As a result, the kagome plane becomes distorted, and the exchange interactions among spins located at the corners of the triangles become spatially nonuniform.  Consequently, the systems magnetically order at low temperature. Cs$_2$Cu$_3$ZrF$_{12}$ undergoes the structural phase transition from $R\bar{3}m$ to the monoclinic space group $P2_1/m$ at 225 K, where the distorted kagome plane is shown in Fig~\ref{fig1}(d), followed by the magnetic phase transition to a N\'eel state at 23.5 K~\cite{Ono:2009hi, Reisinger:2011ij}.  For Cs$_2$Cu$_3$HfF$_{12}$, the structural phase transition occurs at 172 K, and the magnetic transition at 24.5 K~\cite{Ono:2009hi}.  There is no report of the low-temperature-phase crystal structure of Cs$_2$Cu$_3$HfF$_{12}$. 

\begin{figure}
\centering \vspace{0in}
\includegraphics[width=8.5cm]{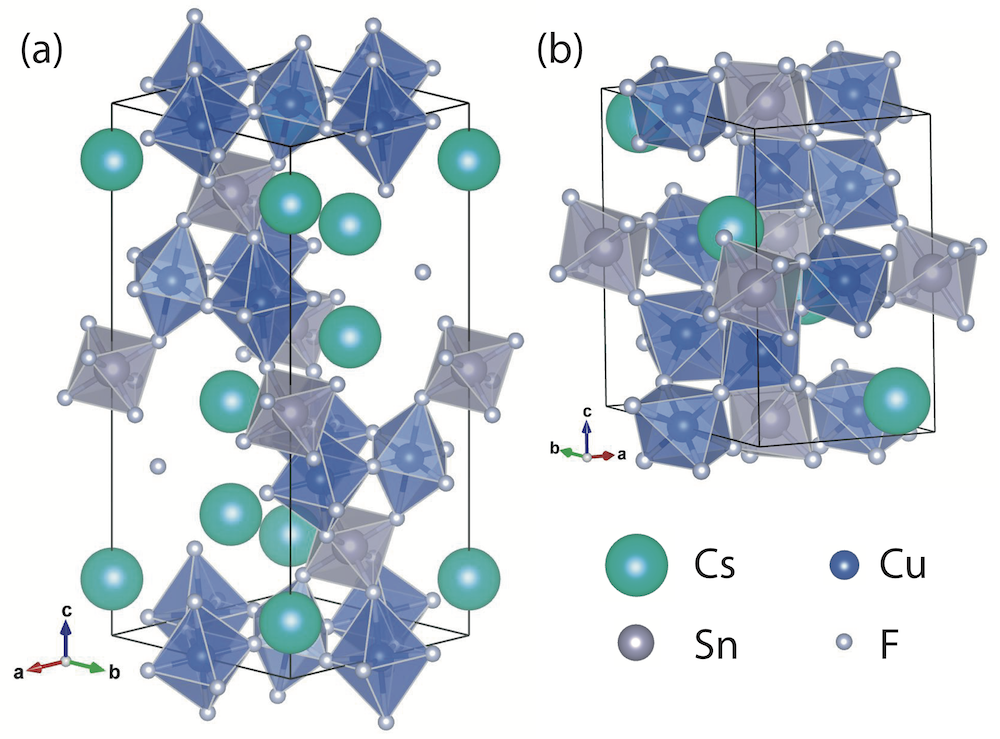}
\caption{(Color online) The crystal structures of Cs$_2$Cu$_3$SnF$_{12}$ in the (a) room-temperature phase and (b) low-temperature phase.}
\label{fig2}
\end{figure}

For Cs$_2$Cu$_3$SnF$_{12}$, at room temperature the compound crystallizes in the rhombohedral space group $R\bar{3}m$ with the lattice parameters $a=7.142(4)$, $c=20.381(14)$ as shown in Fig.~\ref{fig2}(a)~\cite{Ono:2009hi}.  The Cu$^{2+}$ ions in this room-temperature phase form a perfect kagome plane consisting of corner-sharing equilateral triangles.  However, at 185 K, the system undergoes a structural phase transition.  Using combined synchrotron X-ray powder diffraction and high-resolution time-of-flight (TOF) neutron powder diffraction, Downie {\it et al.}~\cite{Downie:2014jg} reported the low-temperature-phase crystal structure to be monoclinic with the space group $P2_1/n$ as shown in Fig.~\ref{fig2}(b).  The phase transition results in structural distortion of the kagome plane giving rise to three inequivalent Cu-Cu bonds as shown in Fig.~\ref{fig1}(a). At low temperature, because of the combined effect of the nonuniform exchange interaction resulting from the distorted kagome lattice and the anisotropic interactions such as the DM interaction, the $S=1/2$ Cu$^{2+}$ spins magnetically order below $T_\textrm{N}=20.2$~K.  The magnetic susceptibility measured on a single crystal sample with a magnetic field applied along the $c$-axis (of the rhombohedral unit cell) shows a cusp at $T_\textrm{N}$~\cite{Ono:2009hi}, and the magnetic order parameter obtained by measuring the scattering intensity of the magnetic Bragg reflection displays a sudden increase at $T_\textrm{N}$~\cite{Ono:2014kp}.   The magnetization suggests that the spins order antiferromagnetically with a small in-plane ferromagnetic component most likely due to small spin canting. A fit of the high-temperature data to the result from the exact diagonalization calculations for the $S=1/2$ uniform Heisenberg kagome antiferromagnet yields the exchange interaction $J/k_\textrm{B}$ of 240~K, attesting to the strong antiferromagnetic interaction.  The ratio of $J/k_\textrm{B}T_\textrm{N}\sim12$ indicates a large degree of frustration. Currently, the magnetic structure has not yet been determined for any magnetically ordered compounds in the $A_2$Cu$_3B$F$_{12}$ family.  

In this article, we report the magnetic structure of Cs$_2$Cu$_3$SnF$_{12}$ measured by high-resolution TOF neutron powder diffraction.  In addition, the magnetic properties are also studied using the magnetization measurements up to an applied magnetic field of 170 T.  The article is organized as follows.  The experiments are discussed in Section II.  The nuclear structures above and below $T_\textrm{t}=185$~K will be presented in IIIA, the high-field magnetization in IIIB, and the magnetic structure below $T_\textrm{N}$ in IIIC. The article ends with the conclusion in Section IV.

\section{Experiments}
Polycrystalline and single crystal samples of Cs$_2$Cu$_3$SnF$_{12}$  were synthesized using the method described in Ref.~\onlinecite{Ono:2009hi}. High-resolution powder neutron diffraction was measured using the TOF neutron diffractometer Super High Resolution Powder Diffractometer (SuperHRPD) at the Japan Proton Accelerator Research Complex (J-PARC).   The sample was loaded into a vanadium can with a diameter of 1 cm, which was cooled to a base temperature of 10 K using a closed-cycle $^4$He cryostat.   The scattering intensity was measured using three sets of detectors, the low-resolution, low-angle detector bank (LA) for the low $\mathbf{Q}$-range, the 90-degree detector bank (QA) for the intermediate $\mathbf{Q}$-range, and the back-scattering detector bank (BS) for high resolution measurements. The powder neutron diffraction measurements were performed at 200 K, 150 K, 25 K, and 10 K.  We note that the structural phase transition from the high-temperature rhombohedral phase to the low temperature monoclinic  phase occurs at $T_\textrm{t}=185$ K and the transition to the magnetically ordered state occurs at $T_\textrm{N}=20.2$ K.  Therefore, at 200 K the system is in the high-temperature phase, in which the crystal structure has been reported to be in the space group $R\bar{3}m$, whereas at 150 K and below, the system is in the low temperature phase of the monoclinic space group $P2_1/n$.  At 10~K, additional magnetic scattering intensity from the magnetically ordered state is also expected.  At each temperature, we obtained three sets of data corresponding to three sets of detectors labelled LA, QA and BS.  The refinement of the diffraction data was done using the Rietveld method implemented in \texttt{FullProf}~\cite{RODRIGUEZCARVAJAL199355}.  In order to determine the structure of the high- and low-temperature phase, global refinement for the 200 K, 150 K, and 25 K data was performed on all three sets of the data simultaneously, where we gave more weight to the high-resolution data taken on the BS detectors than those taken on the LA and QA detectors.  The weighting ratio we use for BS~:~QA~:~LA is $0.90:0.05:0.05$.  In order to determine the magnetic structure of the ordered state, we performed the refinement on the low-$\mathbf{Q}$ region of the 10~K data, where magnetic Bragg peaks were observed.  We put more weight on the data taken using the QA detectors since the measuring range of the QA detectors covers more magnetic Bragg peaks than that of the BS detectors and the resolution of the QA detectors is better than that of the LA detectors; the weighting ratio for BS~:~QA~:~LA is $0.20:0.60:0.20$.  The details of the refinement will be discussed further in the next section.

Additional neutron scattering was performed on the triple-axis spectrometer GPTAS at JRR-3, Japan Atomic Energy Agency, Japan to measure an order parameter related to the structural transition. The neutron diffraction measurements were performed on a single crystal sample using neutron wavelength $k_\textrm{i}=2.676$~\AA$^{-1}$ in a two-axis mode (without analyzing the energy of the scattered neutrons). Collimations of $40'-40'-$sample$-40'$  were employed and two pyrolytic-graphite filters were placed before and after the sample to eliminate higher-order contamination.

In order to obtain information of the spin structure of Cs$_2$Cu$_3$SnF$_{12}$ from the magnitude of the uniform magnetization, the magnetization as a function of the applied magnetic field below $T_\text{N}$ was measured for magnetic fields $B$ parallel and perpendicular to the $c$-axis of the high-temperature rhombohedral phase. The low-field magnetization up to 7\,T was measured using a SQUID magnetometer (Quantum Design MPMS XL). High-field magnetization measurements up to 47~T  were performed using an induction method with a multilayer non-destructive pulse magnet (NDPM) at the International MegaGauss Science Laboratory (IMGSL), Institute for Solid State Physics, University of Tokyo. For the NDPM measurements, we co-aligned crystals, the total weight of which is approximately 100\,mg. For the measurement of $B \perp c$, the crystals were randomly aligned with their $ab$ planes parallel to the magnetic field. 

Since the magnitude of the nearest neighbor exchange interaction of Cs$_2$Cu$_3$SnF$_{12}$ is $J/k_\text{B} = 240$\,K, the critical field at which the 1/3 magnetization ramp~\cite{Nakano2010} or plateau~\cite{Nishimoto2013} occurs is estimated to be $0.85 J \sim 120$\,T. Thus, to observe the magnetization anomaly, we conducted Faraday rotation measurements under an ultra-high magnetic field also at IMGSL. The ultra-high magnetic field was generated using a horizontal single-turn coil megagauss generator~\cite{Herlach1973, Nakao1985}. The maximum field generated was 170\,T for a duration of $\sim 7\,\mu$s. The setup of the optical system for Faraday rotation measurement is detailed in Ref.~\onlinecite{Miyata2011}. To lower the sample temperature below $T_\text{N}$, we used a hand-made liquid-helium flow type cryostat, which is described in Ref.~\onlinecite{Miyata2012}. The Cs$_2$Cu$_3$SnF$_{12}$ sample was shaped to be a thin plate with the wide plane parallel to the cleavage $ab$ plane. The diameter and thickness of the crystal used was 2\,mm and 0.3\,mm, respectively. The sample was placed in the cryostat with the $c$-axis parallel to the magnetic field direction.  A linearly polarized light beam was incident parallel to the $c$-axis. 

\section{Results and discussion}
\subsection{Nuclear structures above and below $T_\textrm{t}$}
\begin{figure*}
\centering \vspace{0in}
\includegraphics[width=17cm]{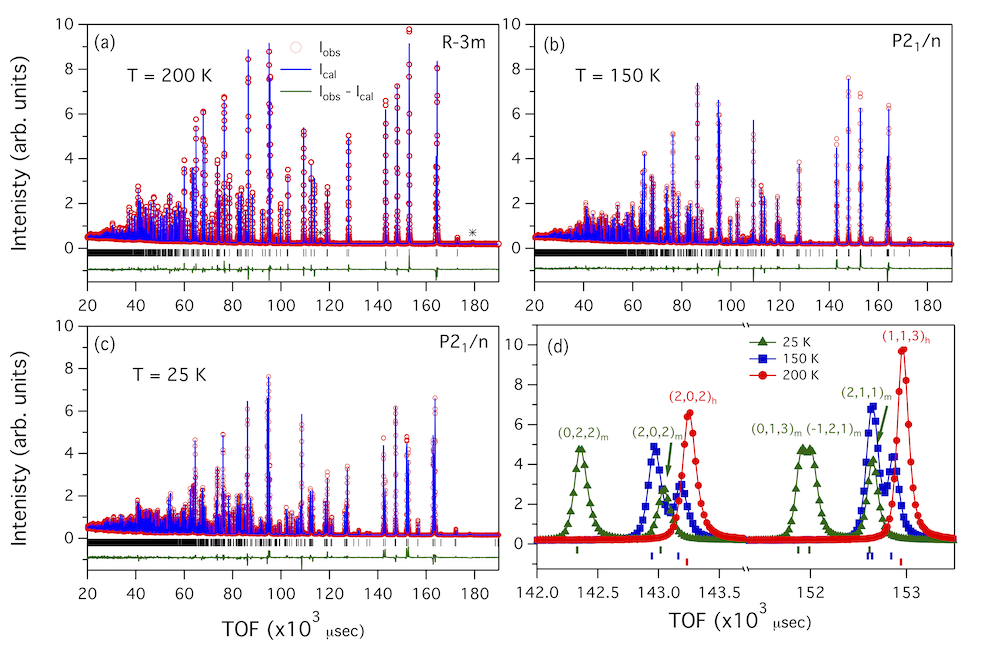}
\caption{(Color online) The TOF powder diffraction was measured on Cs$_2$Cu$_3$SnF$_{12}$ using the BS detectors of SuperHRPD at (a) 200~K, (b) 150~K, and (c) 25~K.  The splitting of the reflection $(1, 1, 3)_\textrm{h}$ and $(2,0,2)_\textrm{h}$ shown in (d) indicates the structural phase transition from the rhombohedral space group $R\bar{3}m$ at 200~K to the monoclinic space group $P2_1/n$ at 150~K and 25~K.  The asterisks denote the impurity phase of Cs$_2$SnF$_6$.}
\label{fig3}
\end{figure*}

Above $T_\textrm{t}=185$~K, the system crystalizes in the rhombohedral crystal system, space group $R\bar{3}m$, where the Cu$^{2+}$ ions form a perfect kagome plane.  Figure~\ref{fig3}(a) shows the result of the refinement of the diffraction data measured by the BS detectors at 200 K to the space group $R\bar{3}m$.  The fit lattice parameters are $a=7.127524(4)~$\AA, $c=20.29099(2)$~\AA, which are slightly lower than those measured at room temperature~\cite{Ono:2009hi, Downie:2014jg} as the unit cell becomes smaller with decreasing temperature.  The refined values of the fractional atomic coordinates are given in Table~\ref{table1}.  We observed a small trace of nonmagnetic Cs$_2$SnF$_6$ as impurity (shown by the asterisk in Fig.~\ref{fig3}(a)).    

\begin{table}
\caption{\label{table1} Refined values of fractional coordinates of Cs$_{2}$Cu$_{3}$SnF$_{12}$ for space group $R\bar{3}m$ at 200 K.}
\centering
\begin{tabular}{c c c c}
\hline \hline
Atom & $x/a$ & $y/b$ & $z/c$\\
\hline
Cs & 0 & 0 & 0.10595(4)\\
Sn & 0 & 0 & 0.5\\
Cu & 0.5 & 0 & 0\\
F(1) &~~~0.20414(3)~~~ &~~~$-0.20414(3)$~~~&~~~0.98427(2)~~~\\
F(2)& 0.13071(3) &$-0.13071(3)$ & 0.44642(2)\\
\hline
\multicolumn{4}{c}{{\sl R$_{1}$} = 0.0308, {\sl wR$_{2}$} = 0.0423}\\
\hline
\end{tabular}
\end{table}

\begin{figure}
\centering \vspace{0in}
\includegraphics[width=9cm]{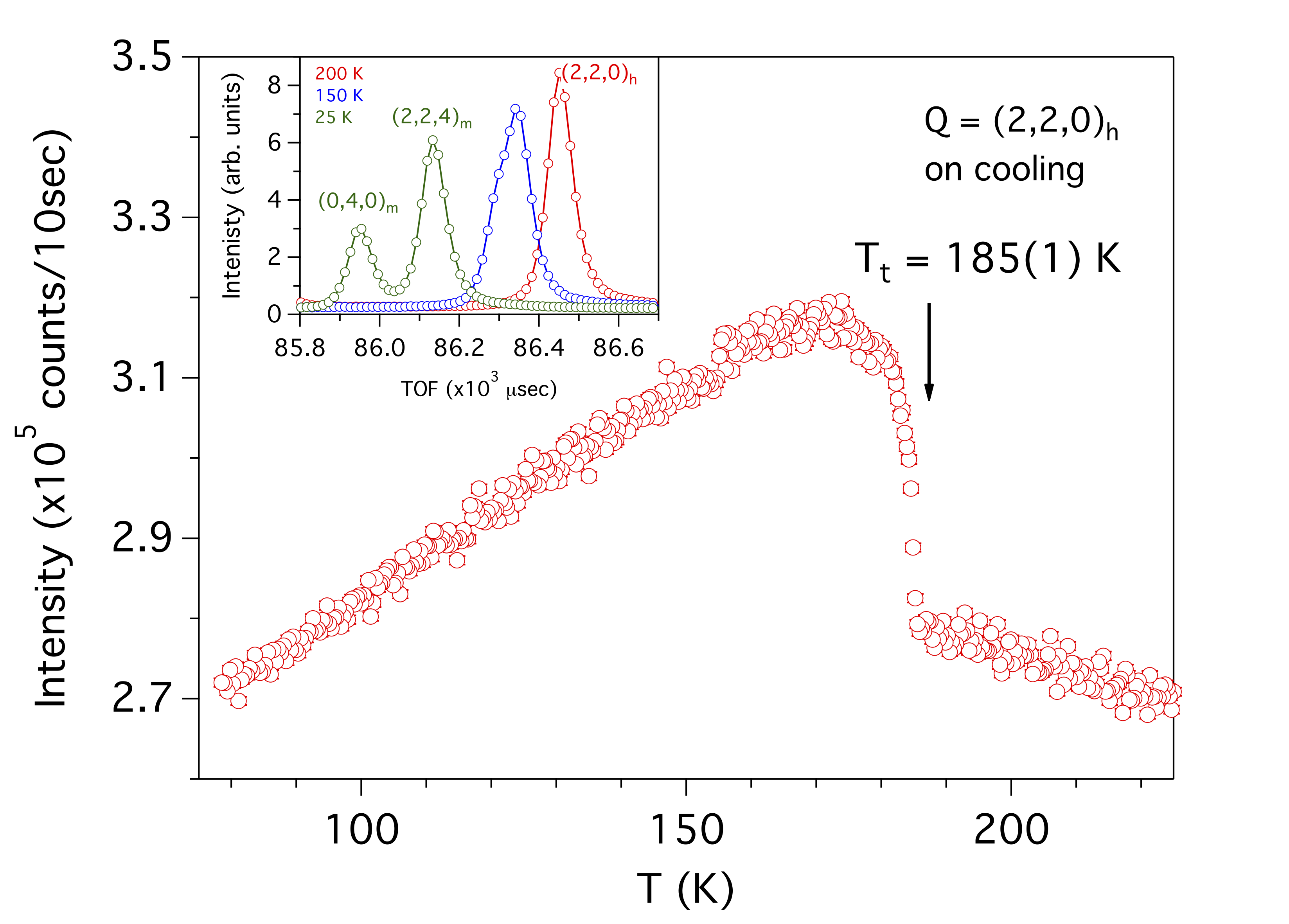}
\caption{(Color online) The scattering intensity of the $(2,2,0)$ reflection (in the high-temperature rhombohedral phase) measured on a single crystal sample using GPTAS shows a sudden increase at 185(1)~K, indicative of the structural transition.  The TOF powder neutron diffraction data (inset) show the splitting of $(2,2,0)$ below $T_\textrm{t}$ to $(2,2,4)$ and (0,4,0) in the low-temperature monoclinic phase.}
\label{fig4}
\end{figure}

At $T_\textrm{t}=185$~K, the system undergoes a structural phase transition to the monoclinic crystal system, space group $P2_1/n$.  Figure~\ref{fig4} shows the neutron scattering intensity measured at the $(2,2,0)_\textrm{h}$ reflection of the rhombohedral high-temperature phase.  The intensity displays an abrupt increase at $T_\textrm{t}=185(1)$~K, indicative of the structural phase transition where $(2,2,0)_\textrm{h}$ splits into $(2,2,4)_\textrm{m}$ and $(0,4,0)_\textrm{m}$ of the monoclinic low-temperature phase (the subscripts $h$ and $m$ denote the rhombohedral and monoclinic crystal systems, respectively).  The previously reported magnetic susceptibility measured on a single crystal synthesized by our group also shows the small anomaly at 185~K in agreement with the order parameter of the scattering intensity reported here. However, this phase transition temperature is significantly higher than that reported in Ref.~\onlinecite{Downie:2014jg}. The difference could be attributed to impurities and defects since in a powder form the compound can easily react with moisture in air.   The sudden increase in the intensity at $T_\textrm{t}$ could be attributed to the splitting of the nuclear Bragg peaks and resolution effect.  The peak broadening at $T_\textrm{t}$ due to the structural transition could result in the observed intensity increase in the triple-axis spectrometer. However, below $T_\textrm{t}$ with decreasing temperature, the split peaks become further apart, and hence the intensity monotonically decreases.

\begin{table}
\caption{\label{table2} Refined values of the fractional coordinates of Cs$_{2}$Cu$_{3}$SnF$_{12}$ for the space group $P21/n$ at 150 K and 25 K.}
\centering
\begin{tabular}{c c c c}
\hline \hline
Atom & $x/a$ & $y/b$ & $z/c$\\
\hline
\multicolumn{4}{c}{T = 150 K}\\
\hline
Cs&~~0.21265(13)~~&~~$-$0.0017(2)~~&~~0.89274(10)~~ \\
 Sn &    0 &  0 &  0.5 \\
 ~~Cu(1)~~ &    0.5 & 0 & 0.5 \\
 Cu(2)  &   0.24820(13) &  0.26143(12)  & 0.25425(8)\\
 F(1) &      0.42477(15) & 0.20429(16)  & 0.39034(12)\\
 F(2) &     0.93915(15)  & 0.68405(16)  & 0.87263(12)\\
 F(3) &      0.23707(13)  & 0.01389(17)  & 0.18756(11)\\
 F(4) &      0.44870(16)  & 0.71466(17)   & 0.10268(12)\\
 F(5) &    $-$0.03104(16)  &  0.17559(17)   & 0.63573(13)\\
 F(6) &      0.76113(13)  & 0.0257(2) & 0.42486(11)\\
\hline
\multicolumn{4}{c}{{\sl R$_{1}$} = 0.0388, {\sl wR$_{2}$} = 0.0431}\\
\hline\hline
\multicolumn{4}{c}{T = 25 K}\\
\hline
 Cs&  0.21377(9) &$-$0.00275(11) & 0.89128(7)\\
 Sn&     0 & 0 & 0.5\\
 Cu(1)&     0.5 &  0 &  0.5\\
 Cu(2)&     0.24637(8) & 0.27036(8) & 0.25757(5)\\
 F(1)&      0.42051(10) & 0.21042(11) & 0.39573(7)\\
 F(2)&      0.94647(10) & 0.67483(11) & 0.86439(8)\\
 F(3)&      0.23816(9) & 0.02504(11) & 0.18493(7)\\
 F(4)&      0.44266(10) &  0.72547(11) & 0.09274(8)\\
 F(5)&~~~$-$0.02380(11)~~~&~~~0.16156(11)~~~&~~~0.64638(8)~~~\\
 F(6)&      0.76115(10) & 0.04323(11) & 0.42874(8)\\
 \hline
\multicolumn{4}{c}{{\sl R$_{1}$} = 0.0376, {\sl wR$_{2}$} = 0.0480}\\
\hline\hline
\end{tabular}
\end{table}

The change in the crystal system from rhombohedral to monoclinic is reflected in the splitting of nuclear  Bragg peaks as previously discussed and illustrated for the $(2,2,0)_\textrm{h}$ reflection (the inset of Fig.~\ref{fig4}). This peak splitting is so small even at 25~K that the intermediate-resolution QA and low-resolution LA detectors are unable to discern the separation among the split peaks.  We can only observe the splitting in the high-resolution BS data.  The structural phase transition from the rhombohedral crystal system to the monoclinic system is in contrast to our previous work~\cite{Ono:2014kp}, in which we reported the structure of the low-temperature phase as having the $2a\times 2a$ enlarged unit cell while retaining the rhombohedral system. This incorrect conclusion in our previous report is due to the lack of the resolution required to detect the peak splitting, highlighting the necessity  of a high-resolution, TOF neutron diffractometer in the refinement of the crystal structure of this compound.  Figures~\ref{fig3}(b) and \ref{fig3}(c) show the fit results of the diffraction intensity measured at 150 K and 25 K, respectively.  The fit lattice parameters at 150~K are $a=7.909309(9)~$\AA, $b=7.114155(9)~$\AA, $c=10.624465(2)~$\AA, and $\beta=97.9840(1)^\circ$,  and at 25~K are $a=7.890704(8)~$\AA, $b=7.085362(7)~$\AA, $c=10.565583(11)~$\AA, and $\beta=97.72622(9)^\circ$.  These lattice parameters appear to be consistent with the previous reported values measured at 100~K, taking into account the temperature difference. The refined values of the fractional atomic coordinates at both temperatures are given in Table~\ref{table2}, and they are also in good agreement with the previous work~\cite{Downie:2014jg}.  

The structural phase transition introduces distortion into the SnF$_6$ and CuF$_6$ octahedra resulting in a distorted kagome plane with three inequivalent bonds as shown in Fig.~\ref{fig1}(a).  The distortion can be quantified by the percentage of deviation in Cu$-$F$-$Cu bond angles from the value measured at 200~K, which at 25~K are 0.3\%, 1.4\%, and 2.6\%.   The splitting of two reflections, $(1,1,3)_\textrm{h}$ and $(2,0,2)_\textrm{h}$, at three different temperatures 200~K (above $T_\textrm{t}$), 150~K and 25~K (below $T_\textrm{t}$) are shown in Fig.~\ref{fig3}(d) to illustrate the evolution of the split peaks as a function of temperature. Downie {\it et al.} suggested that the structural transition in Cs$_2$Cu$_3$SnF$_{12}$ is driven by the tendency of the Sn$^{4+}$ ions to reduce their bond valence sum (BVS)~\cite{Downie:2014jg}.  We performed the BVS calculations using \texttt{BondStr}~\cite{Brown:2237510}.  In agreement with their report, we observed the decrease of BVS for the Sn$^{4+}$ ions surrounded by six F$^-$ ions from 4.604(2) at 200~K to 4.480(7) at 150~K and 4.416(4) at 25~K.  On the other hand, the BVS of the Cu$^{2+}$ ions in the similar octahedral environment of CuF$_6$ remains relatively constant in the same temperature range.  

\subsection{High-field magnetization}
\begin{figure}[htb]
	\centering
	\includegraphics[width=8cm]{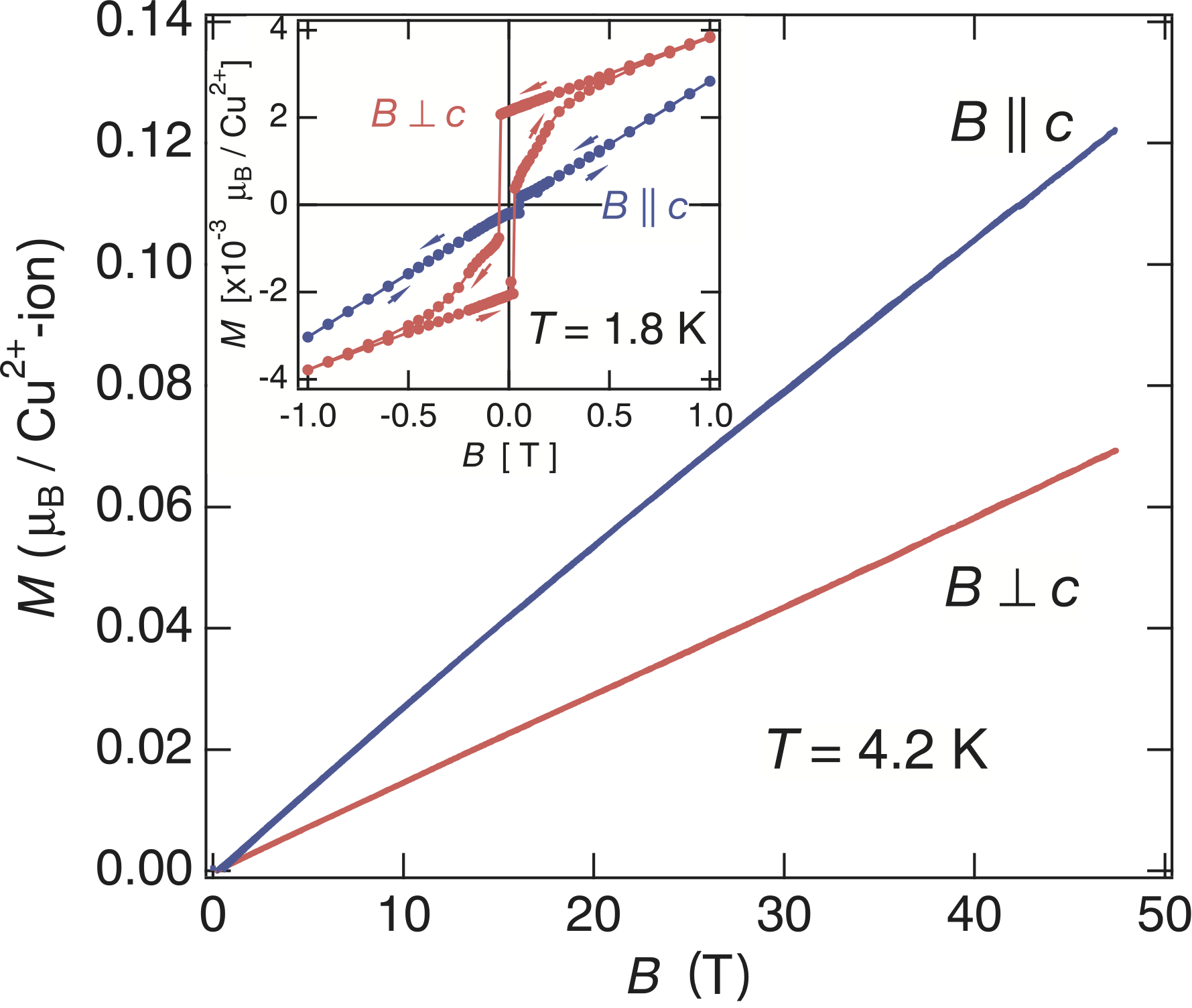}
	\caption{(Color online) Magnetization curves measured at $T=4.2$\,K for $B \parallel c$ (blue symbols) and $B \perp c$ (red symbols) using a non-destructive pulse magnet. The inset shows the hysteresis of the magnetization curves measured at 1.8\,K using the SQUID magnetometer.
}
	\label{fig5}
\end{figure}

The magnetic susceptibility of Cs$_2$Cu$_3$SnF$_{12}$ measured at 1 T as a function of temperature has already been reported in Ref.~\onlinecite{Ono:2009hi}, where the data show the structural phase transition at $T_\textrm{t}$ and the transition to the magnetically ordered state at $T_\textrm{N}$.  In this work, high-field magnetization measurements were performed to investigate possible field-induced magnetic transitions up to the applied field of 170 T.  The main panel of Fig.\,\ref{fig5} shows the magnetic field dependence of the magnetization measured at $T= $~4.2\,K for magnetic fields parallel and perpendicular to the $c$ axis using NDPM. For both field directions, the magnetization increases monotonically while exhibiting slight convex curves up to 47~T without noticeable anomalies. The slope of the magnetization for $B \parallel c$, where the $c$-axis is of the rhombohedral phase, is about 1.7 times larger than that for $B \perp c$. This magnetization anisotropy mainly arises from the anisotropy of the $g$-factors, {\it i.e.}, $g_{\parallel c}=2.48$ and $g_{\perp c}=2.10$, and from the DM interaction~\cite{Ono:2009hi}. The inset shows the hysteresis of the magnetization curves measured at 1.8\,K using the SQUID magnetometer. No spontaneous magnetic moment was detected at zero field for $B \parallel c$ (perpendicular to the kagome plane), while for $B \perp c$ (within the kagome plane) a finite spontaneous magnetic moment of $2.4\times 10^{-3}\,\mu_\text{B}/\text{Cu}^{2+}$ spin was observed. The relationship between the observed magnetic moment and the magnetic structure will be discussed later. 

\begin{figure*}[htb]
	\centering
	\includegraphics[width=17cm]{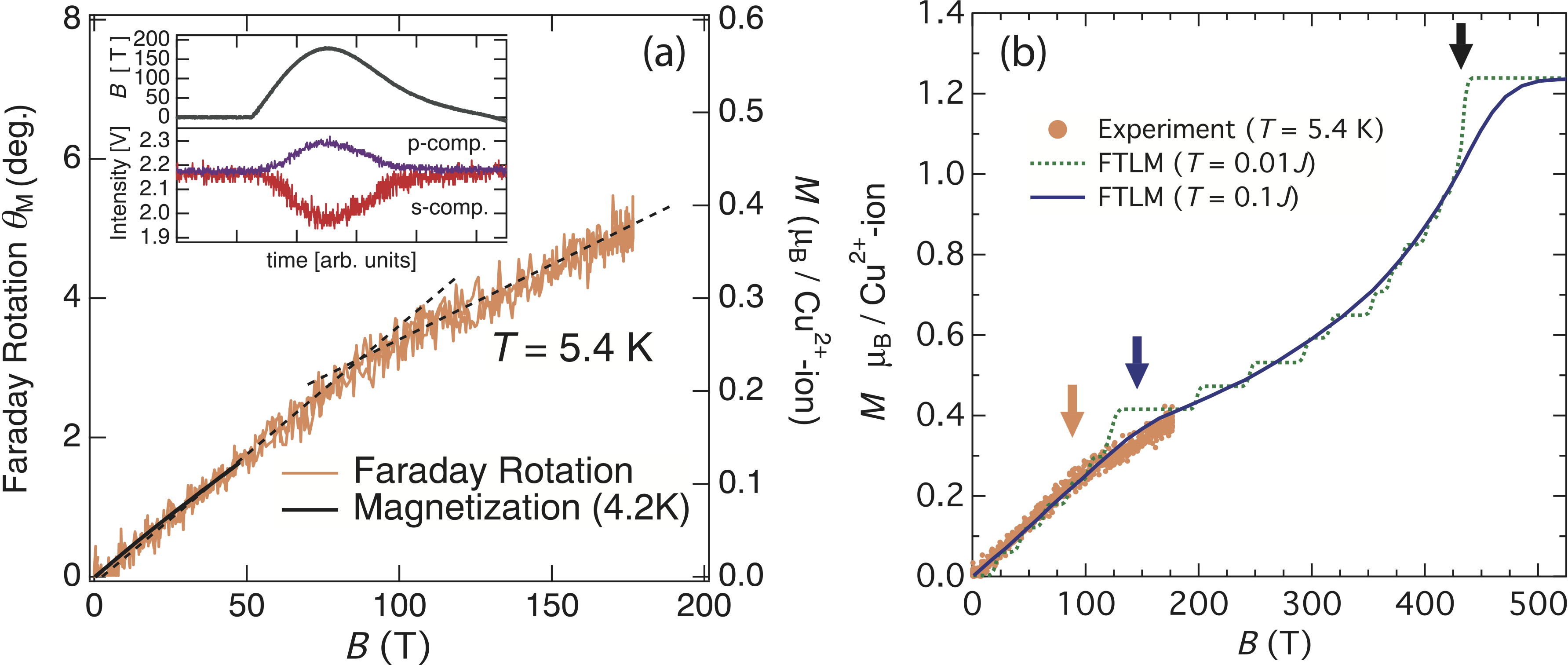}
	\caption{(Color online) (a) The magnetic field dependences of the Faraday rotation angle measured at 5.4\,K  for $B \parallel c$ (orange line) and the magnetization estimated from the Faraday rotation angle. The thick black line indicates the magnetization data obtained using a non-destructive pulse magnet, which was used for the correction of the diamagnetic component $\delta\theta_\text{dia}$ and the calibration of the scale factor between the magnetization and the Faraday rotation angle $\theta_\text{M}$. The dashed lines are the guides representing the magnetization slopes of the low- and high-field regions. (b) The measured magnetization is compared with the results from the FTLM discussed in the text.  The orange and blue arrows denote the bend in the experimental and calculated magnetization, respectively.  The black arrow represents the saturation field $B_\textrm{s}$ of 432 T.}
	\label{fig6}
\end{figure*}

For the ultra-high-magnetic-field measurements up to 170~T, Figure~\ref{fig6} shows the magnetic field dependence of the Faraday rotation angle measured at 5.4\,K for $B \parallel c$ and the magnetization estimated from the Faraday rotation angle. The inset shows the time dependence of magnetic field generated by the single-turn coil megagauss generator (upper panel) and transmitted light intensity through the sample (lower panel). Blue and red lines in the lower panel of the inset denote horizontally ($p$) and vertically ($s$) polarized components of the transmitted light intensity $I_\text{p}$ and $I_\text{s}$, respectively. The Faraday rotation angle $\theta_\text{F}$ is given by 
\[
	\theta_\text{F} = \frac{1}{2}\sin^{-1}\left(\frac{I_\text{p}-I_\text{s}}{I_\text{p}+I_\text{s}}\right).
\] 
The observed $\theta_\text{F}$ includes the diamagnetic component $\delta\theta_\text{dia}$, which arises from the diamagnetism of the sample.  As discussed in Ref.~\onlinecite{Kojima2008}, we subtracted $\delta\theta_\text{dia}$ from $\theta_\text{F}$ so that the resultant Faraday rotation angle $\theta_\text{M}$ reproduces the magnetization $M$ measured up to 47\,T using NDPM, which is shown by the solid black line in Fig.~\ref{fig6}. The orange line in Fig.\,\ref{fig6} is the Faraday rotation angle $\theta_\text{M}$ corrected for the diamagnetic component, $\theta_\text{M} = \theta_\text{F} - \delta\theta_\text{dia}$.  The magnetization increases almost linearly as a function of magnetic field with no prominent  anomaly up to 170~T.  This is suggestive of the robustness of the magnetic structure against the external magnetic field.  However, the magnetization slope changes at around 90\,T, as shown by dashed lines in  Fig.~\ref{fig6}.  The magnetization at the bending point is roughly 0.24$\mu_\textrm{B}/\textrm{Cu$^{2+}$}$. This deviation from a straight line could be suggestive of the phase transition from the low-field 120$^\circ$ spin structure, which will be discussed in the next section, to another spin structure stabilized by the quantum fluctuation.  The theoretical calculation by means of a grand canonical density matrix renormalization group indicates the existence of a series of magnetization plateaus as a function of field at 0, 1/9, 1/3, 5/9, and 7/9 of the saturated magnetization~\cite{Nishimoto2013}.  Whereas the 0 and 1/9 plateaus have the characteristics of a quantum spin liquid state, the 1/3, 5/9, and 7/9 plateaus exhibit the symmetry-breaking long-range order.  Out of the five plateaus, the 1/3 plateau, which results from one up-spin and two spins forming a singlet on a triangular plaquette, appears to be the most dominant, making it the most likely to be experimentally observed~\cite{Nishimoto2013, PhysRevB.88.144416, Hida_JPSJ.70.3673}.  However, we note that since the ground state of Cs$_2$Cu$_3$SnF$_{12}$ is not the quantum spin liquid state, the 1/3 plateau in this system, if proven to exist, might not consist of a singlet pair (later discussed).  Furthermore, most of the calculations of magnetization as a function of applied field were performed at zero temperature, hindering the direct comparison with the experimental data measured at a finite temperature, particularly when the change of the magnetization at a plateau is subtle.  

In an attempt to explain the change of slope in the magnetization data measured on Cs$_2$Cu$_3$SnF$_{12}$, we compare the data with the large-scale ($N=42$) numerical calculations using the finite-temperature Lanczos method (FTLM) performed by Schnack {\it et al.}~\cite{PhysRevB.98.094423}. Figure~\ref{fig6}(b) shows qualitative similarity between the FTLM result at a finite temperature ($T = 0.1J$) shown by the blue solid line and the experimental data with the appearance of the knee-like bend instead of the step-like plateau, which is smoothened out due to thermal effects.  However, the calculated values of the 1/3 magnetization, which is estimated to be 0.41~$\mu_\textrm{B}$, and of the critical field of about 145 T (indicated by the blue arrow in Fig.~\ref{fig6}(b)) are significantly different from the observed values of 0.24~$\mu_\textrm{B}$ and 90 T (indicated by the orange arrow), respectively.  We note that the magnetization of the FTLM result was renormalized by the saturated moments of $g_c\mu_\textrm{B} S=1.24~\mu_\textrm{B}$, where $g_c=2.48$ and $S=1/2$, and the applied field was rescaled by the saturation field calculated from $g_c\mu_\textrm{B} B_\textrm{s}=3J$~\cite{Nishimoto2013,PhysRevB.88.144416, Hida_JPSJ.70.3673}, where $J/k_\textrm{B}=240$~K, yielding $B_\textrm{s}=432$~T.  The discrepancy could be ascribed to the effect of the anisotropic terms in the spin Hamiltonian and spatially nonuniform exchange interactions.  

A similar bend in the magnetization, indicative of the 1/3 magnetization plateau, was experimentally observed in the $S=1/2$ distorted kagome antiferromagnets Rb$_2$NaTi$_3$F$_{12}$, Cs$_2$NaTi$_3$F$_{12}$, and Cs$_2$KTi$_3$F$_{12}$~\cite{PhysRevB.94.104432}, the $S=1$ distorted kagome antiferromagnets Cs$_2$KV$_3$F$_{12}$, Cs$_2$NaV$_3$F$_{12}$, and Rb$_2$NaV$_3$F$_{12}$~\cite{PhysRevB.95.134436}, the $S=3/2$ undistorted kagome-lattice antiferromagnet KCr$_3$(OH)$_6$(SO$_4$)$_2$~\cite{JPSJ.80.063703}, and the distorted kagome antiferromagnets Cs$_2$KCr$_3$F$_{12}$ and Cs$_2$NaCr$_3$F$_{12}$~\cite{PhysRevB.97.224421}.  In all of these observations, the step-like 1/3 magnetization plateau appears as a kink or knee-like bend due to thermal effects.  More recent work on Cd-kapellasite (CdCu$_3$(OH)$_6$(NO$_3$)$_2\cdot$H$_2$O), comprising antiferromagnetically coupled Cu$^{2+}$ $S=1/2$ spins on a perfect kagome network, shows a series of magnetization plateaus, including the 1/3 plateau, which are interpreted as crystallized magnons localized on the hexagon of the kagome lattice~\cite{Okuma:2019dr}.  As with Cs$_2$Cu$_3$SnF$_{12}$, Cd-kapellasite magnetically orders at low temperature ($T_\textrm{N}=4$~K) to the 120$^\circ$ structure with weak in-plane ferromagnetism resulted from DM-interaction-induced spin canting~\cite{Okuma:2017dx}.  Hence, it is possible that the plateau-like bend observed at 90 T in Cs$_2$Cu$_3$SnF$_{12}$ might have the same origin, which is different from the previously discussed theoretical model proposed in Ref.~\onlinecite{Nishimoto2013}.  Unfortunately, the exchange coupling for Cs$_2$Cu$_3$SnF$_{12}$ is about 5 times larger than that of Cd-kapellasite ($J/k_\textrm{B}=45$~K giving the calculated saturation field $B_\textrm{s}\sim100$~T and the measured $B_\textrm{s}$ of $\sim160$~T), making the saturated magnetization state unreachable with the current most powerful magnet [Fig.~\ref{fig6}(b)].  We therefore cannot conclusively state the origin of the anomaly at 90~T, and hence more detailed measurements up to a much higher field are required.

\subsection{Magnetic structure below $T_\textrm{N}$}

\begin{figure*}
	\begin{center}
		\includegraphics[width=17cm]{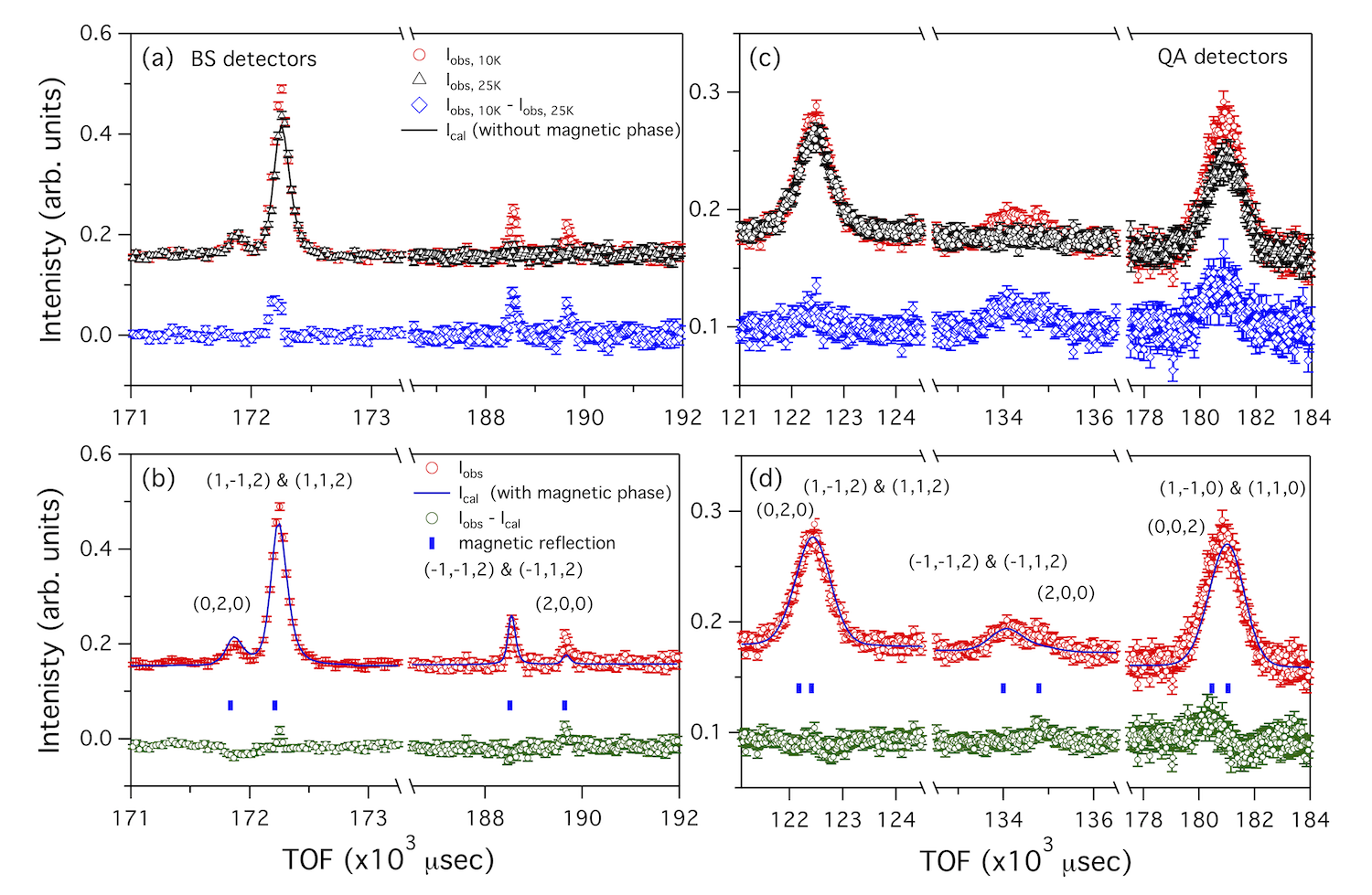}
		\caption{(Color online) The TOF powder neutron diffraction measured at 10~K using the (a) BS detectors and (b) QA detectors shows extra intensity, indicative of magnetic scattering, when compared to that at 25~K.  (c) and (d) display fits to calculated intensity based on the all-in-all-out structure ($P2_1'/n'$ with positive vector chirality) for the BS data and QA data, respectively.}
		\label{fig7}
	\end{center}
\end{figure*}

Below $T_\textrm{N}=20.2$~K, the Cu$^{2+}$ spins in Cs$_2$Cu$_3$SnF$_{12}$ magnetically order.  The TOF powder neutron diffraction reveals a larger scattering intensity for the low-$Q$ (large TOF) reflections at 10~K than at 25~K.  Figures~\ref{fig7}(a) and \ref{fig7}(c) show the extra scattering intensity at 10 K, which was measured by the BS and QA detector, respectively.  This extra scattering shown by the blue data points in Fig.~\ref{fig7} cannot be accounted for by the nuclear scattering intensity shown by the solid black lines, which represent the results from the refinement performed at 25 K, and is attributed to the magnetic Bragg scattering.  The appearance of the magnetic scattering on top of the crystallographic nuclear reflection indicates the magnetic ordering vector $\mathbf{k}=(0,0,0)$, which will become useful in the discussion of the magnetic space group analysis below.  For the BS detector, we can clearly observe only three magnetic Bragg peaks at $(2,0,0)$, $(-1,-1,2)$\&$(-1,1,2)$ and $(1,-1,2)$\&$(1,1,2)$; all reflections measured at low temperatures are labelled using the monoclinic system and the subscript $m$ will henceforth be ignored.

The magnetic Cu$^{2+}$ ions occupy two distinct Wyckoff positions, the high-symmetry position $2c$ for Cu(1) with the multiplicity of 2 and general low-symmetry position $4e$ for Cu(2) with the multiplicity of 4.  The irreducible representation ({\it irrep}) analysis was employed for both Cu positions using  \texttt{BasIreps}~\cite{Hovestreydt:wi0099} with $P2_{1}/n$ as the input for the underlying crystallographic structure. The resulting {\it irreps} for both Wyckoff positions are shown in Table~\ref{IRCu}.  The decompositions of the magnetic representations for Cu(1) and Cu(2) are
\begin{align}
\Gamma(2c) &= 3\Gamma^1\oplus0\Gamma^2\oplus3\Gamma^3\oplus0\Gamma^4\\
\Gamma(4e) &= 3\Gamma^1\oplus3\Gamma^2\oplus3\Gamma^3\oplus3\Gamma^4,
\end{align}
where all {\it irreps} are of one dimension.  Since the magnetic moments at both sites order together, they must order under the same {\it irreps}, $\Gamma^1$ or $\Gamma^3$.  We note that for Cu(1) with 2 spins (Cu(2) with 4 spins) there are 6 (12) free parameters for the magnetic moments.  However, in the magnetically ordered state, only one {\it irrep} will be selected.  Hence, by using the {\it irrep} analysis the number of free parameters for each Cu$^{2+}$ spins is decreased to 3.  Later, we will assume that the magnitude of the magnetic moments at 2 sites are the same, which further lowers the number of free parameters.  In order to determine the {\it irrep} of the ordered state and its magnetic space group, we will next consider the symmetry invariance of the magnetic structure, which is derived from the underlying crystallographic space group. 

\begin{table*}
	\caption{\label{IRCu}(Color online) Irreducible representations ({\it irreps}) and their basis vectors for Cu(1) at Wyckoff positions $2c$, Cu(1)$_1$ at (0.5, 0.0, 0.5) and Cu(1)$_2$ at (0.0, 0.5, 0.0) and for Cu(2) at Wyckoff positions $4e$, Cu(2)$_1$ at (0.2464, 0.2704, 0.2576), Cu(2)$_2$ at (0.2536, 0.7704, 0.2424), Cu(2)$_3$ at (0.7536, 0.7296, 0.7424), and Cu(2)$_4$ (0.7464, 0.2296, 0.7576). $\Gamma^2$ and $\Gamma^4$ are not admissible for Cu(1) due to the symmetry of the $2c$ positions.  The highlighted space group is the magnetic space group for the ordered state of Cs$_2$Cu$_3$SnF$_{12}$.}	
	\begin{tabular}{c | c | c || c | c | c | c | c}
		\hline \hline
		~~~{\it irrep}~~~  & {Cu(1)$_1$} & {Cu(1)$_2$} & {Cu(2)$_1$} & {Cu(2)$_2$} & {Cu(2)$_3$} & {Cu(2)$_4$} &~Shubnikov group~ \\
		\hline
		$\Gamma^1$ &~$(u_1, v_1, w_1)$~&~$(-u_1, v_1, -w_1)$~&~$(u_2, v_2, w_2)$~&~$(-u_2, v_2, -w_2)$~&~$(u_2, v_2, w_2)$~&~$(-u_2, v_2, -w_2)$~& $P2_1/n$ \\ \hline
		$\Gamma^2$ & $-$ & $-$ &~$(u_2, v_2, w_2)$~&~$(-u_2, v_2, -w_2)$~&~$(-u_2, -v_2, -w_2)$~&~$(u_2, -v_2, w_2)$~& $P2_1/n'$ \\ \hline
		{\color{blue}$\Gamma^3$} &~$(u_1, v_1, w_1)$~&~$(u_1, -v_1, w_1)$~&~$(u_2, v_2, w_2)$~&~$(u_2, -v_2, w_2)$~&~$(u_2, v_2, w_2)$~&~$(u_2, -v_2, w_2)$~& {\color{blue}$P2_1'/n'$}  \\ \hline
		$\Gamma^4$ & $-$ & $-$ &~$(u_2, v_2, w_2)$~&~$(u_2, -v_2, w_2)$~&~$(-u_2, -v_2, -w_2)$~&~$(-u_2, v_2, -w_2)$~& $P2_1'/n$ \\ 
		\hline \hline	
	\end{tabular}
	
      	\vspace{1ex}
	\small
	\raggedright Note that for $k=(0,0,0)$, $(u_1, v_1, w_1)$, and $(u_2, v_2, w_2)$ can be viewed as the magnetic moments along the non-orthogonal crystallographic axes of space group $P2_1/n$ for Cu(1) and Cu(2), respectively. 
\end{table*}

The point group for the space group $P2_{1}/n$ of the underlying crystal structure is $2/m$ of order 4, which consists of 4 elements: $\mathbf{G}=\{1, 2_y, m_y, \bar{1}\}$.  This point group is abelian, and hence all aforementioned {\it irreps} are of one dimension. From this point group, the black-white magnetic point group $\mathbf{M}$ can be obtained using the relation $\mathbf{M}=\mathbf{H}+(\mathbf{G}-\mathbf{H})1'$, where $\mathbf{H}$ are the subgroups of index 2 of $\mathbf{G}$ and $1'$ is the spin reversal (time-reversal) operator.  For $\mathbf{G}=2/m$, $\mathbf{H}$ can be $\{1, 2_y\}$, $\{1, m_y\}$, and $\{1, \bar{1}\}$.   Hence, the magnetic point groups for point group $P2_1/n$ are the single-color point group (Type I) $2/m$, the grey point group (Type II) $2/m1'$, and three black-white point groups (Type III) $2/m'$, $2'/m$, and $2'/m'$~\cite{Izyumov:2012ud}.  

Based on the crystallographic space group $P2_1/n$, similar symmetry analysis for space groups gives rise to 5 possible magnetic space groups or Shubnikov groups, excluding the black-white space groups of the second kind (Type IV).  We note that in the table of magnetic space groups~\cite{Litvin}, these groups are listed based on $P2_1/c$; however one can simply change $P2_1/c$ to $P2_1/n$ by changing a unit cell (by convention, the axes are chosen so that $\beta$ is closer to 90$^\circ$).  The 5 magnetic space groups are (1) $P2_1/n$, a monochrome (single-color) space group (Type I) that is the same as the underlying crystallographic space group, (2) $P2_1/n1'$, a grey space group (Type II) that can be used to describe a magnetic system in a paramagnetic state (not applicable to the magnetic ordered state), (3) $P2_1'/n$, (4) $P2_1/n'$, and (5) $P2_1'/n'$; the last three Shubnikov groups are called the black-white space groups of the first kind (Type III), which can be used to describe the magnetic ordered state that has the same unit cell as the crystallographic counterpart. We will not consider the black-white space groups of the second kind (Type IV), which describes the magnetic structure with the enlarged magnetic unit cell, since the magnetic unit cell of Cs$_2$Cu$_3$SnF$_{12}$ is the same as the crystallographic unit cell; the magnetic ordering vector is $\mathbf{k}=(0,0,0)$ and hence the magnetic order is of the $q=0$ type.  

Out of the 4 magnetic space groups that can be used to describe the magnetically ordered state, only two magnetic space groups, $P2_1/n$ and $P2_1'/n'$, are admissible for the magnetic structure of Cs$_2$Cu$_3$SnF$_{12}$ since due to symmetry, $P2_1'/n$ and $P2_1/n'$ yield a zero magnetic moment for the Wyckoff positions $2c$ of Cu(1), which is also magnetic.  In terms of {\it irreps}, $P2_1/n$ corresponds to {\it irrep} $\Gamma_1$ and $P2_1'/n'$ to {\it irrep} $\Gamma_3$ as shown in Table~\ref{IRCu}.  Therefore, the {\it irrep} and space-group symmetry analysis lead to two possible candidates, namely $P2_1/n$ and $P2_1'/n'$, for the magnetic space group of the magnetically ordered phase in Cs$_2$Cu$_3$SnF$_{12}$.  The analysis significantly decreases the number of free parameters from 18 parameters for 6 spins in a unit cell to 6 parameters for Cu(1) and Cu(2) spins.  Because of the paucity of observed magnetic Bragg reflections and their weak intensity combined with the fact that they appear on top of the structural peaks,  the refinement would become difficult or impossible without the symmetry analysis, attesting to its significance to the magnetic structure refinement of a complex system with weak scattering intensity. 

Because of the weak magnetic scattering intensity as indicated by Fig.~\ref{fig7}, we will also have to rely on the result of the magnetization measurements to put a constraint on the magnetic structure and reduce the number of free parameters even further.  As previously discussed, the magnetization shows that the three spins at the corners of each triangle most likely form a structure that slightly deviates from the coplanar 120$^\circ$ structure, resulting in a small but finite in-plane ferromagnetic moment.   This in-plane ferromagnetic moment is ascribed to spin canting, and has a value per Cu$^{2+}$ spin at zero field, extracted from the magnetization, of $2.4\times10^{-3}$ $\mu_\textrm{B}$.  This small canted moment would be difficult to resolve using the available neutron scattering data, but it does give a constraint on the magnetic structure as will be later discussed.  Using this constraint in combination with the fact that the magnitude of the magnetic moment at the two copper sites must be the same, we can reduce the number of the free parameters in the magnetic structure refinement. 

\begin{figure}
\centering \vspace{0in}
\includegraphics[width=8.5cm]{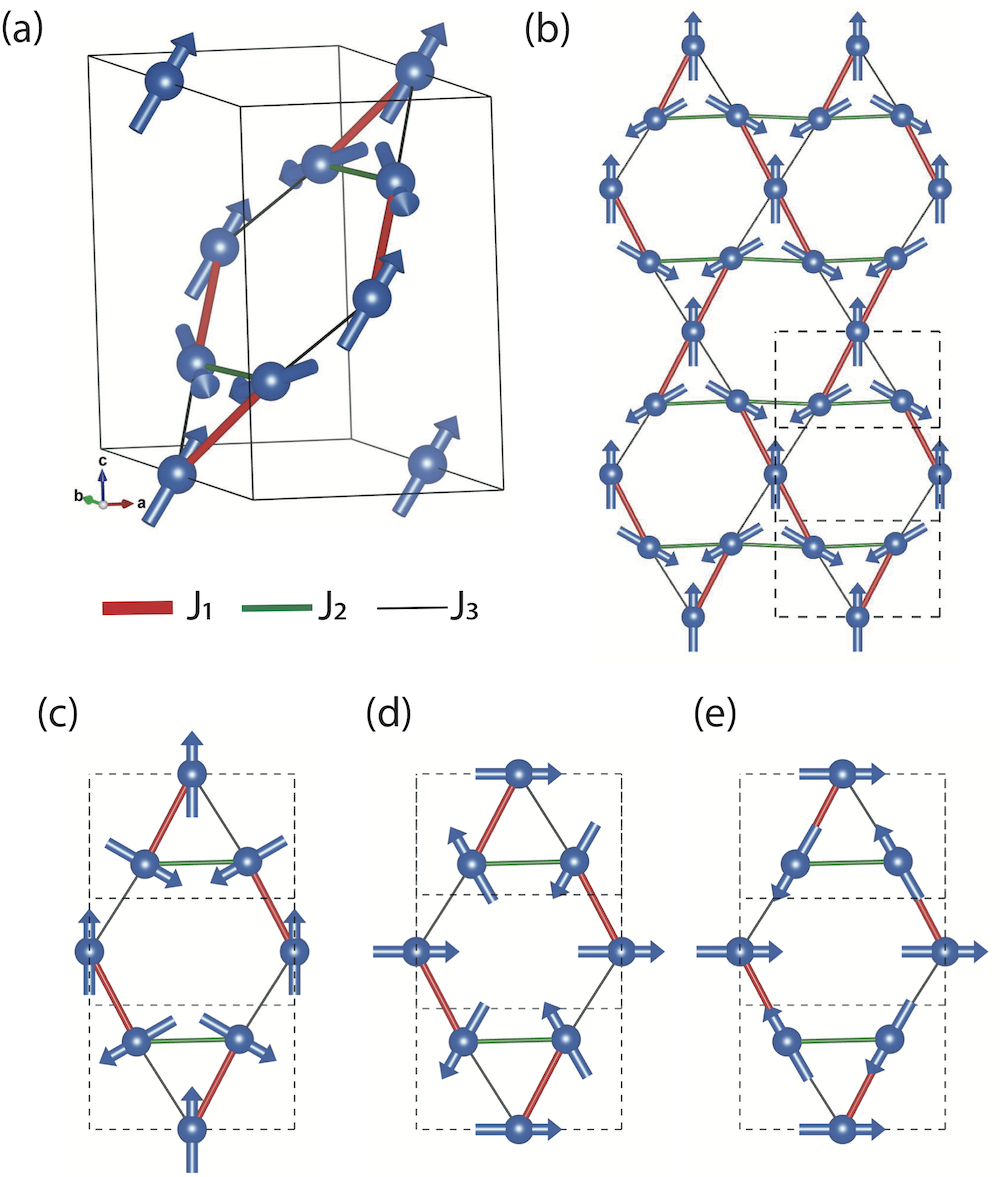}
\caption{(Color online) The all-in-all-out spin ($P2_1'/n'$ with positive vector chirality) structure of Cs$_2$Cu$_3$SnF$_{12}$ is shown (a) within the unit cell of space group $P2_1/n$, and (b) in the kagome plane.  The dashed line in (b) denotes the unit cell projected onto the kagome plane. (c), (d), and (e) illustrate the spin structures corresponding to $P2_1'/n'$ with negative vector chirality, $P2_1/n$ with positive vector chirality, and $P2_1/n$ with negative vector chirality, respectively.}
\label{fig8}
\end{figure}

\begin{table*}
	\caption{\label{refined-const}(Color online) Refined magnetic moments in a units of $\mu_\textrm{B}$ of Cu(1) (0.5, 0.0, 0.5) and Cu(2) (0.2464, 0.2704, 0.2576) for magnetic space groups $P2_1'/n'$ and $P2_1'/n'$ with positive and negative vector chirality.  }
	\centering
	\begin{tabular}{ c | c | c | c | c }
		\hline \hline
		\multicolumn{1}{c|}{\multirow{1}{*}{~~~~~~magnetic space group~~~~~~}}& \multicolumn{1}{c|}{\multirow{1}{*}{~~~~~vector chirality~~~~~}} & \multicolumn{1}{c|}{\multirow{1}{*}{~~~~~$|M|~(\mu_\textrm{B})$~~~~~}} &  \multicolumn{2}{c}{~~~~~~Magnetic $R$-factor\footnote{The magnetic R-factors are for the QA data, which are given the most weight in the magnetic structure refinement.} (\%)~~~~~~} \\
		 \multicolumn{1}{c|}{} &\multicolumn{1}{c|}{}& \multicolumn{1}{c|}{} &~~~~~Cu(1)~~~~~&~~~Cu(2)~~~\\
		 \hline
		 {\color{blue}$P2_1'/n'$} & {\color{blue}positive} &~~{\color{blue}$0.68(3)$}~~~ & {\color{blue}6.05} & {\color{blue}7.85}\\
		  & negative &~~$0.67(3)$~~~ & 6.17 & 7.99\\
		\hline
		 $P2_1/n$ & positive &~~$0.54(2)$~~~ & 11.4 & 13.0 \\
		  & negative &~~$0.57(3)$~~~ & 11.6 & 13.1\\
		\hline \hline
	\end{tabular}
	
\end{table*}

Using the combined results from the {\it irrep} (magnetic space group) analysis and indirect evidence from the magnetization discussed above, we performed the refinement of the magnetic structure on the 10 K data taken on all three detectors simultaneously. We assume that the spins lie on the kagome plane and form the 120$^\circ$ spin structure with net zero moment. To first approximation, we will ignore the small in-plane canted moment.  A coplanar magnetic structure is possible for both magnetic space groups $P2_1/n$ and $P2_1'/n'$.  In order to refine the magnetic structure, the 25~K data were first fit to the crystallographic space group $P2_{1}/n$ to obtain the crystal structure, atomic parameters, and peak profile parameters; this task has been discussed in the previous section.  We then take the fit results from the refinement of the 25 K data, and first fit the 10 K data without the magnetic phase.  For the magnetic structure refinement, since the magnetic scattering is observed at low $\mathbf{Q}$ due to the magnetic form factor, the refinement was performed for $|\mathbf{Q}| < |\mathbf{Q}_\textrm{c}| \sim 1.9$~\AA$^{-1}$.  For the BS, QA, and LA detectors, $|\mathbf{Q}_\textrm{c}|$ corresponds to the cutoff time-of-flight TOF$_\textrm{c}$ of 160.0 ms, 113.0 ms, and 41.2 ms, respectively.  We note that no extra magnetic scattering intensity for Bragg peaks with $|\mathbf{Q}| > |\mathbf{Q}_\textrm{c}$ (TOF $<$ TOF$_\textrm{c}$) was visibly observed.  All parameters related to the nuclear structure of the system were then fixed throughout the magnetic structure refinement, and only spin parameters, $i.e.$, magnetic moments or directions, $\theta$ and $\phi$, were fit.  However, from the above discussion, we fix the spin directions so that the three spins on the corners of the triangle form the 120$^\circ$ structure.  We found that the atomic positions and atomic displacement parameters do not significantly change between 25~K and 10~K.  The lattice parameters change slightly from 25~K to 10~K but this change is significant enough to affect the peak positions and hence warrant the refitting at 10~K.  The solid lines in Figs.~\ref{fig7}(a) and~\ref{fig7}(c) denote the fit without the magnetic phase, whereas the fit with the magnetic phase is represented by the lines in Fig.~\ref{fig7}(b) for the BS data and Fig.~\ref{fig7}(d) for the QA data.

\begin{figure}
\centering \vspace{0in}
\includegraphics[width=7cm]{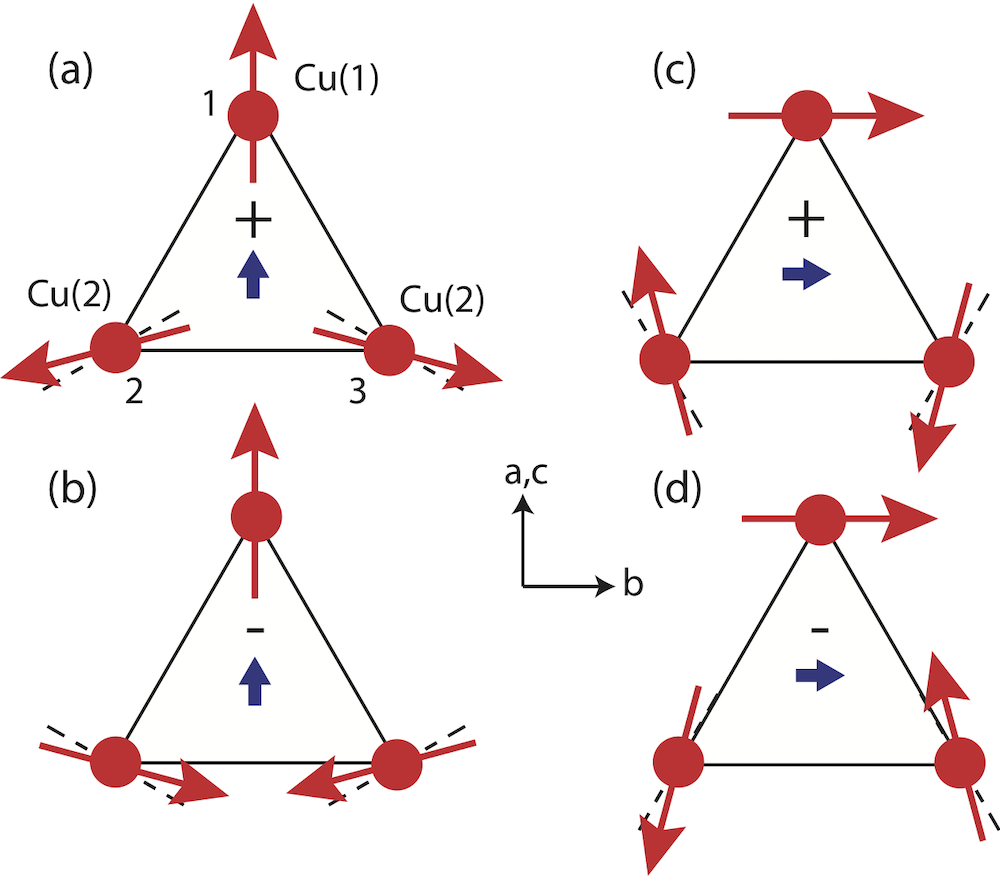}
\caption{(Color online) Possible canting spin structures allowed by symmetry of magnetic space groups are shown for (a) $P2_1'/n'$ with positive vector chirality, (b) $P2_1'/n'$ with negative vector chirality, (c) $P2_1/n$ with positive vector chirality, and (d) $P2_1/n$ with negative vector chirality.  The canted moment for $P2_1'/n'$ shown in (a) and (c) [$P2_1/n$ shown in (b) and (d)] is perpendicular (parallel) to the monoclinic $b$-axis.  For clarity, the canting is exaggerated and does not depict the actual canting, which is much smaller. The $+$ and $-$ signs denote the positive and negative vector chirality, whereas the thick blue arrow on each triangle represents the direction of the in-plane canted moment.}
\label{fig9}
\end{figure}

For the magnetic space group $P2_1'/n'$, the refinement based on the 120$^\circ$ structure with positive vector chirality (the all-in-all-out spin structure) as shown in Figs.~\ref{fig7}(b) and \ref{fig7}(d) with the magnitude of the spin moment as the only fitting parameter yields the magnetic moment of 0.68(3)$\mu_{B}$ with the global $\chi^{2} = 5.84$; for comparison, excluding the magnetic structure refinement, the global $\chi^2$ becomes 8.72.  The spin structure for the all-in-all-out structure is shown in Fig.~\ref{fig8}(a) for the spins in the monoclinic unit cell and in Fig.~\ref{fig8}(b) for the spins forming the kagome plane.  The vector chirality of each triangle for the coplanar 120$^\circ$ structure is defined as~\cite{Grohol:2005bg}
\[
\mathbf{K}=\frac{2}{3\sqrt{3}}\left(\mathbf{\hat{S}}_1\times\mathbf{\hat{S}}_2+\mathbf{\hat{S}}_2\times\mathbf{\hat{S}}_3+\mathbf{\hat{S}}_3\times\mathbf{\hat{S}}_1\right),
\]
where $\mathbf{\hat{S}}_i$ with $i=1,2,3$ are the unit vectors representing the directions of the spins around the triangle as shown in Fig.~\ref{fig9}(a).  $\mathbf{K}$ is perpendicular to the plane with amplitude $+1$ (positive vector chirality) or $-1$ (negative vector chirality).  We note that the spin structure where the two spins on the base of the triangle switch, changing the vector chirality of the spins around the triangle from positive to negative shown in Fig.~\ref{fig8}(c), is also consistent with magnetic space group $P2_1'/n'$ but gives a slightly higher value of the magnetic $R$-factor than that for the all-in-all-out structure (Table~\ref{refined-const}). 

For the magnetic space group  $P2_1/n$, all spins rotate $90^\circ$ within the kagome plane with respect to the spin structure in $P2_1'/n'$ as shown in Figs.~\ref{fig8}(d) and \ref{fig8}(e) for the positive and negative vector chirality, respectively.  The fit to the magnetic space group $P2_1/n$ yields a worse result than the fit to $P2_1'/n'$ does as evidenced by the significantly larger magnetic $R$-factors listed in Table~\ref{refined-const}. Hence, out of the four considered 120$^\circ$ spin structures, the all-in-all-out structure ($P2_1'/n'$ with positive vector chirality) gives the best fit to the neutron diffraction data.  However, the sign of the vector chirality cannot be conclusively determined due to the small difference in the magnetic R-factors for the structures with positive and negative vector chirality.  According to the symmetry of the magnetic space groups, the in-plane canting is allowed for both magnetic space groups $P2_1/n$ and $P2_1'/n'$.  However, the allowed canted moments for the two magnetic space groups are expected to be orthogonal to each other as illustrated by Fig.~\ref{fig9}.  Hence, in order to confirm the all-in-all-out magnetic structure, in-plane magnetization measurements with varying the direction of the applied magnetic field are required.  These measurements are in fact being prepared and will be reported elsewhere.  We note that even though the coplanar 120$^\circ$ spin structure is assumed as suggested by the magnetization data (previously discussed), based on our neutron diffraction data, we cannot rule out the possibility of a finite out-of-plane magnetic component, which is also allowed for the magnetic space group $P2_1'/n'$. Other techniques such as NMR, which can probe the local magnetic field generated by spin magnetic moments, can potentially provide valuable complementary data to refine the magnetic structure of Cs$_2$Cu$_3$SnF$_{12}$; concurrently, the NMR measurements have been performed and the result will be published elsewhere~\cite{mihael}.

The fit value of the ordered moment of 0.68(3)~$\mu_\textrm{B}$ for $P2_1'/n'$ with positive vector chirality is about 32\% smaller than the expected value of 1 $\mu_\textrm{B}$ for the $S=1/2$ spin.  A similar reduction of the ordered moment was also observed in jaroiste, KFe$_3$(OH)$_6$(SO$_4$)$_2$, which has the same all-in-all-out spin structure.  In jarosite, the measured ordered moment is 3.80(6)~$\mu_\textrm{B}$, which is 24\% smaller than the expected value of 5~$\mu_\textrm{B}$ for the $S=5/2$ Fe$^{3+}$ spin~\cite{Inami.PhysRevB.61.12181}.  The reduced ordered moment might result from geometric frustration of the underlying triangular-based spin network and quantum fluctuations, given that we observed the larger percentage reduction in Cs$_2$Cu$_3$SnF$_{12}$ with $S=1/2$ than in jarosite with $S=5/2$.

For the $A_2$Cu$_3B$F$_{12}$ family, a ground state appears to be strongly influenced by the underlying spin network and hence the crystal structure.  For Cs$_2$Cu$_3$SnF$_{12}$, the low-temperature crystal structure gives rise to the spin network shown in Fig.~\ref{fig1}(a) with three distinct exchange interactions.  We note that this spin network is different from the one (shown in Fig.~\ref{fig1}(c)) that was used in the spin-wave calculations in Ref.~\onlinecite{Ono:2014kp}.  However, this difference does not significantly change the average value of the effective exchange interactions, and hence does not affect the main finding reported in Ref.~\onlinecite{Ono:2014kp}.  The magnitudes of these interactions can be estimated by the Cu$-$F$-$Cu bond angles~\cite{Goodenough, Kanamori}, which are 139.48(5)$^\circ$, 137.93(5)$^\circ$, and 136.31(5)$^\circ$ for $J_1>J_2>J_3$, respectively. Compared to other systems in this family such as Rb$_2$Cu$_3$SnF$_{12}$ with four distinct exchange interactions, where the Cu$-$F$-$Cu angles range from 124$^\circ$ for the weakest bond to 138$^\circ$ for the strongest bond, the values of the nonuniform exchange interactions in  Cs$_2$Cu$_3$SnF$_{12}$ are expected to be close to one another.  Presumably, the system is closer to the uniform kagome antiferromagnet than Rb$_2$Cu$_3$SnF$_{12}$.   For Rb$_2$Cu$_3$SnF$_{12}$, the strongly nonuniform exchange interactions with the strong $J_1$ bonds that form a pinwheel pattern give rise to the pinwheel valence bond solid state that appears to be robust against the small anisotropic DM interaction.  On the other hand, for Cs$_2$Cu$_3$SnF$_{12}$, the more uniform exchange interaction and the different pattern of the spin network renders the system more vulnerable (being close to the ideal case) to small perturbations in the spin Hamiltonian such as the next-nearest-neighbor and DM interactions; as a result, Cs$_2$Cu$_3$SnF$_{12}$ magnetically orders at low temperatures, and the magnetic structure of the ordered state is most likely determined by the small contribution from the DM interaction~\cite{Elhaja.PhysRevB.66.014422}.  The small in-plane ferromagnetism could result from the combined effect of the DM interaction and the spatially nonuniform exchange interactions. Compared to Cs$_2$Cu$_3$ZrF$_{12}$ and Cs$_2$Cu$_3$HfF$_{12}$, the magnetization measured in these two compounds shows a much larger anomaly at $T_\textrm{t}$ than that observed in Cs$_2$Cu$_3$SnF$_{12}$~\cite{Ono:2009hi}, suggesting that the structural transition could lead to larger lattice distortion and hence larger deviation of the exchange interactions from the uniform case.   One would therefore expect that the spin structures for Cs$_2$Cu$_3$ZrF$_{12}$ and Cs$_2$Cu$_3$HfF$_{12}$ could be different from the 120$^\circ$ structure.  The magnetic structures, which are currently unknown, of these two compounds could substantiate this observation. 

\section{Conclusion}
High-resolution TOF neutron diffraction measurements on Cs$_2$Cu$_3$SnF$_{12}$ confirm the structural phase transition from the rhombohedral space group $R\bar{3}m$ to the monoclinic space group $P2_1/n$ at $T_\textrm{t}=185$~K.  As a result of the lattice distortion in the low-temperature structural phase, the exchange interactions become spatially nonuniform. The $S=1/2$ Cu$^{2+}$ spins magnetically order below $T_\textrm{N}=20.2$~K.  Combined magnetization and neutron powder diffraction data below $T_\textrm{N}$ suggest that the ordered state  has the coplanar 120$^\circ$ spin structure in the magnetic space group $P2_1'/n'$ with positive vector chirality (the all-in-all-out structure).  The magnetic ordering could be attributed to the DM interaction and the spatially nonuniform exchange interactions. The magnetization measurements at low fields reveal ferromagnetism within the kagome plane.  The magnetic space group analysis for $P2_1'/n'$ confirms that the in-plane canting is allowed by symmetry.  However, the in-plane direction of the weak ferromagnetic moment cannot be resolved using the neutron diffraction data.  The high-field magnetization measured in the applied field up to 170 T varies linearly as a function of field and has a the subtle knee-like bend (a change of slope) around 90 T, which could be indicative of the 1/3 magnetization plateau and the transition from the classical 120$^\circ$ N\'eel state to an interesting field-induced magnetic state. Further work is required to elucidate this high-field state.

\begin{acknowledgments}
Work at Mahidol University was supported in part by the Thailand Research Fund Grant Number RSA6180081 and the Thailand Center of Excellence in Physics (ThEP). The neutron diffraction experiments were approved by the Neutron Science Proposal Review Committee of J-PARC MLF (Proposal No. 2012B0234 for SuperHRPD) and supported by the Inter-University Research Program of KEK.  The ``JSPS Invitational Fellowships for Research in Japan'' supported the stay of KM at IMRAM, Tohoku University, where some part of this work was carried out.  HT was supported by Grants-in-Aid for Scientific Research (A) (Grant Nos.~17H01142, 23244072 and 26247058) from the Japan Society for the Promotion of Science.  KM would like to thank M. A. Allen for useful suggestions. \\[3mm]
\end{acknowledgments}

\clearpage
\onecolumngrid
\begin{center}\rule{\textwidth}{0.8pt}\end{center}
\section*{Erratum (2026)}
\noindent\emph{The following Erratum, correcting the magnetic structure reported above, has been
published as Phys.~Rev.~B \textbf{114}, 139902(E) (2026).  Within it, ``the original paper'' refers to the
article reproduced above; cross-references to its figures and tables have been adapted to the
numbering of the present combined document, and its reference list has been merged with that of
the article.  No other change has been made to the accepted text.}
\begin{center}\rule{\textwidth}{0.8pt}\end{center}
\twocolumngrid

In the original paper the ordered moments of the two crystallographically inequivalent Cu$^{2+}$ sites were placed in two \emph{separate} magnetic phases, so that their magnetic Bragg intensities were added incoherently instead of coherently, as a single $\mathbf{k}=0$ magnetic structure requires.  Refinements with the correct coherent summation favor the coplanar $120^{\circ}$ structure of magnetic space group $P2_1/n$ with negative vector chirality and ordered moment $0.52(2)\,\mu_{\mathrm{B}}$, rather than the all-in-all-out structure of $P2_1'/n'$ with $0.68(3)\,\mu_{\mathrm{B}}$ reported originally; that identification is therefore withdrawn.  The corrected structure reverses the sign of the out-of-plane Dzyaloshinskii--Moriya component, forbids a net out-of-plane moment by symmetry, and agrees with recent nuclear quadrupole resonance (NQR) results~\cite{Grbic2026}.  All other results of the original paper are unaffected.

Prompted by a recent NQR study by Grbi\'c et al.~\cite{Grbic2026}, which finds a spin structure consistent with $P2_1/n$ and negative vector chirality, we re-evaluated our previous neutron powder diffraction analysis, and discovered an error.  In the original paper, the magnetic structure of Cs$_2$Cu$_3$SnF$_{12}$ at 10~K was refined against the time-of-flight neutron powder diffraction data measured on SuperHRPD (J-PARC) using the {\scshape FullProf} suite.  In that analysis, the ordered moments on the two crystallographically inequivalent Cu$^{2+}$ sites, Cu(1) at $(0.5, 0.0, 0.5)$ and Cu(2) at $(0.2464, 0.2704, 0.2576)$, were described by two \emph{separate} magnetic phases.  Since the calculated pattern of a multiphase model is the sum of the intensities of the constituent phases, the magnetic Bragg intensities of the two sublattices were added incoherently, $I(\mathbf{Q}) \propto |\mathbf{F}_{\!\perp}^{\mathrm{Cu(1)}}(\mathbf{Q})|^{2}+|\mathbf{F}_{\!\perp}^{\mathrm{Cu(2)}}(\mathbf{Q})|^{2}$, where $\mathbf{F}_{\!\perp}$ denotes the component of the magnetic structure factor perpendicular to the scattering vector $\mathbf{Q}$.  This treatment is incorrect.  The two sites order within a single magnetic structure with propagation vector $\mathbf{k}=0$, and their magnetic structure factors must therefore be summed coherently,
\begin{equation}
I(\mathbf{Q}) \;\propto\; \bigl|\mathbf{F}_{\!\perp}^{\mathrm{Cu(1)}}(\mathbf{Q})+\mathbf{F}_{\!\perp}^{\mathrm{Cu(2)}}(\mathbf{Q})\bigr|^{2},
\label{eq:erratum-coherent}
\end{equation}
which differs from the incoherent sum by the interference term $2\,\mathrm{Re}\,[\mathbf{F}_{\!\perp}^{\mathrm{Cu(1)}}\!\cdot\!\mathbf{F}_{\!\perp}^{\mathrm{Cu(2)}*}]$.  This term is not negligible: for the all-in-all-out structure it suppresses, for example, the calculated intensity of the $(0,1,1)$ magnetic reflection by a factor of about five relative to the incoherent sum.

\begin{table*}[t]
	\caption{\label{tab:erratum}Corrected refinement results for the four coplanar $120^{\circ}$ spin structures, replacing Table~\ref{refined-const}.  The two Cu sublattices are now included in a single magnetic phase [Eq.~(\ref{eq:erratum-coherent})], with the common moment magnitude $|M|$ as the only refined magnetic parameter.  The magnetic $R$ factor ($R_{\mathrm{mag}}$, integrated-intensity Bragg $R$ factor of the magnetic phase) is listed for each of the three data banks: backscattering (BS), $90^{\circ}$ (QA, given the most weight in the refinement), and low-angle (LA); the global $\chi^{2}$ (Bragg contribution) is for the three data banks refined together.  All values were obtained with {\scshape FullProf.2k} version 7.95 (2023), whereas version 6.20 (2018) was used in the original paper.  With version 7.95 the \emph{original} (incoherent) two-phase all-in-all-out model yields the global $\chi^{2}=5.45$; global $\chi^{2}$ values should therefore be compared only within a single program version.  Boldface highlights the best-fitting structure and its magnetic space group.}
	\begin{ruledtabular}
	\begin{tabular}{lcccccc}
		\multicolumn{1}{l}{\multirow{2}{*}{Magnetic space group}} & \multicolumn{1}{c}{\multirow{2}{*}{Vector chirality}} & \multicolumn{1}{c}{\multirow{2}{*}{$|M|~(\mu_{\mathrm{B}})$}} & \multicolumn{3}{c}{$R_{\mathrm{mag}}$ (\%)} & \multicolumn{1}{c}{\multirow{2}{*}{global $\chi^{2}$}} \\
		 & & & BS & QA & LA & \\
		\hline
		$P2_1'/n'$ & Positive & $0.60(2)$ & 14.5 & 14.9 & 42.5 & 5.56 \\
		 & Negative & $0.48(2)$ & 24.4 & 11.5 & 32.5 & 5.57 \\
		$\bm{P2_1/n}$ & Positive & $0.40(1)$ & 20.2 & 14.4 & 28.1 & 5.41 \\
		 & \textbf{Negative} & $\bm{0.52(2)}$ & $\bm{4.9}$ & $\bm{5.2}$ & $\bm{29.1}$ & $\bm{5.21}$ \\
	\end{tabular}
	\end{ruledtabular}
\end{table*}

\begin{figure*}[t]
	\includegraphics[width=\textwidth]{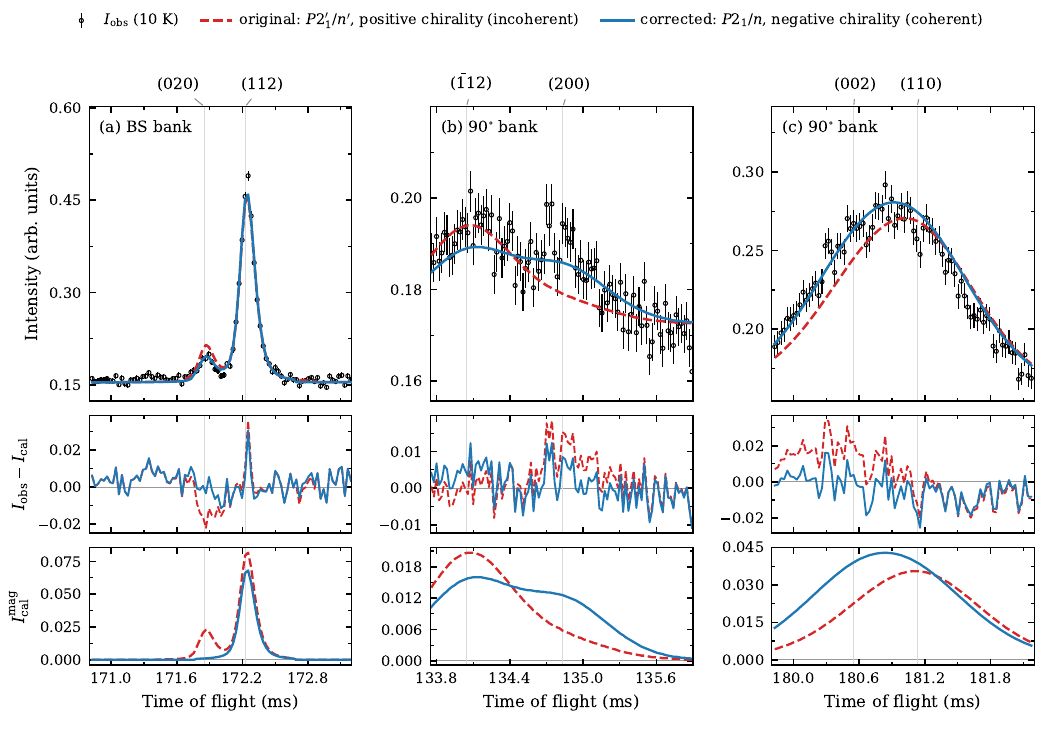}
	\caption{\label{fig:erratum}Original and corrected fits to the 10~K neutron powder diffraction data, shown for the magnetic reflections discussed in the text.  (a) Backscattering bank, where $(0,2,0)$ and $(1,1,2)$ are resolved; (b),(c) $90^{\circ}$ bank around $(\bar{1},1,2)/(2,0,0)$ and $(0,0,2)/(1,1,0)$.  Vertical grey lines mark the calculated reflection positions.  Top row: observed intensity (circles, with statistical error bars) together with the calculated profile of the original refinement (red dashed; all-in-all-out structure of $P2_1'/n'$, the two Cu sublattices in separate magnetic phases and therefore added incoherently) and of the corrected refinement (blue solid; $P2_1/n$ with negative vector chirality, both sublattices in a single magnetic phase and added coherently).  Middle row: the corresponding difference curves $I_{\mathrm{obs}}-I_{\mathrm{cal}}$, plotted on a common vertical scale within each panel.  Bottom row: the calculated magnetic contribution alone for the two models.  The original model places substantial magnetic intensity at $(0,2,0)$, where the data show none, and too little at $(2,0,0)$ and $(0,0,2)$; the corrected model removes both discrepancies.}
\end{figure*}

We have repeated the magnetic structure refinements with the moments on both Cu sites included in a \emph{single} magnetic phase, so that Eq.~(\ref{eq:erratum-coherent}) is satisfied.  All other aspects of the analysis are identical to those of the original paper: the same magnetic symmetry operators of the magnetic space groups $P2_1'/n'$ and $P2_1/n$, the same three data banks (backscattering, $90^{\circ}$, and low-angle, weighted 0.2:0.6:0.2), the same fixed ideal $120^{\circ}$ moment directions [Figs.~\ref{fig8}(a)--\ref{fig8}(e)], and the magnitude of the ordered moment as the only refined magnetic parameter.  The corrected results are summarized in Table~\ref{tab:erratum}, which replaces Table~\ref{refined-const}.  Because the two sublattices now belong to one magnetic phase, a single magnetic $R$ factor per data bank is obtained for each model.

The ordered moment is revised accordingly.  If the all-in-all-out structure ($P2_1'/n'$ with positive vector chirality) is imposed, the coherent refinement yields $0.60(2)\,\mu_{\mathrm{B}}$ with the global $\chi^{2}=5.56$, instead of $0.68(3)\,\mu_{\mathrm{B}}$ with the global $\chi^{2}=5.84$ reported in the original paper; for the best-fitting structure (see below) the ordered moment is $0.52(2)\,\mu_{\mathrm{B}}$.  The moment reduction from the expected $1\,\mu_{\mathrm{B}}$ for the $S=1/2$ spin is therefore $40$--$48\%$, rather than the $32\%$ quoted originally.

The correction also reverses the model preference expressed by the neutron data.  In the original (incoherent) analysis the refinements based on $P2_1'/n'$ gave distinctly smaller magnetic $R$ factors ($6$--$8\%$) than those based on $P2_1/n$ ($11$--$13\%$), which was the basis for identifying the all-in-all-out structure.  Under the correct coherent treatment the best fit is instead obtained for \emph{$P2_1/n$ with negative vector chirality} [Fig.~\ref{fig8}(e)], with the magnetic $R$ factor $5.2\%$ (QA bank) and the global $\chi^{2}=5.21$, compared with $14.9\%$ and $5.56$ for the all-in-all-out structure (Table~\ref{tab:erratum}); the negative-chirality $P2_1/n$ structure gives the smaller magnetic $R$ factor in each of the three data banks, so this preference is unchanged under any reweighting of them.  The preference has a transparent origin.  For $P2_1/n$ the magnetic intensity of the $(0,2,0)$ reflection vanishes identically: at $(0,k,0)$ the $a$ and $c^{*}$ moment components of the screw-related sublattices cancel exactly, and the surviving $b$ component is parallel to the scattering vector.  For $P2_1'/n'$, by contrast, $(0,2,0)$ is among the strongest calculated magnetic reflections.  The 10~K data show essentially no magnetic intensity at $(0,2,0)$ beyond the nuclear peak, and the intensity distribution over the $(1,1,0)$, $(0,0,2)$, $(1,1,2)$, and $(2,0,0)$ reflections likewise follows the $P2_1/n$ negative-chirality pattern.  The two fits are compared directly for these reflections in Fig.~\ref{fig:erratum}: the original refinement predicts a magnetic peak at $(0,2,0)$ that the data do not show [Fig.~\ref{fig:erratum}(a)] and it underestimates the observed intensity at $(2,0,0)$ and around $(0,0,2)$ [Figs.~\ref{fig:erratum}(b) and \ref{fig:erratum}(c)], whereas the corrected refinement accounts for all of them.  The weak in-plane ferromagnetism reported in the original paper does not by itself discriminate between the two magnetic space groups: all four candidate structures are fully compensated (zero net moment), and a symmetry-allowed in-plane canting exists for \emph{both} magnetic space groups, with mutually orthogonal in-plane directions [Fig.~\ref{fig9}]; the in-plane direction of the canted moment relative to the crystallographic axes was not determined.  The out-of-plane magnetization, however, does bear on the model choice, as discussed below.

The corrected best-fit structure---the coplanar $120^{\circ}$ structure of $P2_1/n$ [irreducible representation $\Gamma_1$ of the original paper] with negative vector chirality---has two physical implications.  First, it implies a sign change of the component of the Dzyaloshinskii--Moriya (DM) vector perpendicular to the kagome plane, $D_z$.  As discussed in the original paper, the ordered state of the kagome antiferromagnet is selected by the DM interaction, and the sign of $D_z$ determines the vector chirality of the $\mathbf{k}=0$ $120^{\circ}$ structure: one sign stabilizes the positive-chirality (all-in-all-out) state and the opposite sign the negative-chirality state~\cite{Elhaja.PhysRevB.66.014422,Grohol:2005bg}.  The revision of the refined structure from positive to negative vector chirality therefore reverses the sign of $D_z$ inferred for Cs$_2$Cu$_3$SnF$_{12}$; in particular, the sign is now opposite to that in jarosite, where $D_z$ selects the positive-chirality all-in-all-out structure~\cite{Grohol:2005bg,Inami.PhysRevB.61.12181}.  Second, the negative-chirality structure does not allow out-of-plane canted moments.  Whereas for positive vector chirality the in-plane DM component cants the spins out of the kagome plane into the umbrella (all-in-all-out) configuration, for negative vector chirality it induces no out-of-plane canting and the spins remain strictly coplanar~\cite{Elhaja.PhysRevB.66.014422}; equivalently, in the magnetic space group $P2_1/n$ the only symmetry-allowed uniform magnetization is along the $b$ axis, which lies in the kagome plane, so that a net out-of-plane moment is forbidden.  (In $P2_1'/n'$, by contrast, a net out-of-plane component is symmetry allowed, and the original paper noted that it could not be excluded by the neutron data.)  This is in agreement with the magnetization data reported in the original paper: at low field a spontaneous in-plane moment of $2.4\times10^{-3}\,\mu_{\mathrm{B}}$ per Cu$^{2+}$ was observed, while no spontaneous moment was detected perpendicular to the kagome plane.  The absence of an out-of-plane spontaneous moment, which within $P2_1'/n'$ would require the symmetry-allowed out-of-plane canting to be accidentally small, follows by symmetry in the corrected structure, providing independent support for it.

Consequently, the identification of the all-in-all-out structure ($P2_1'/n'$ with positive vector chirality) as the magnetic structure of Cs$_2$Cu$_3$SnF$_{12}$ must be withdrawn.  Although the magnetic Bragg reflections are weak and, because $\mathbf{k}=0$, coincide with nuclear reflections, the pronounced differences in the magnetic $R$ factors identify the coplanar $120^{\circ}$ structure of magnetic space group $P2_1/n$ with negative vector chirality, with ordered moment $0.52(2)\,\mu_{\mathrm{B}}$, as the correct magnetic model.  This determination is corroborated by the absence of an out-of-plane spontaneous moment in the magnetization data, as discussed above, and is in full agreement with the recent NQR findings~\cite{Grbic2026}.  All results concerning the crystal structure, the structural and magnetic transition temperatures, the symmetry analysis of the allowed canting, and the magnetization and high-field measurements are unaffected.

\emph{Disclaimer.} An artificial-intelligence assistant (Claude Fable, Anthropic) was used to reanalyze the neutron powder diffraction data, including the identification of the incoherent-summation error, the corrected coherent {\scshape FullProf} refinements, and the symmetry analysis reported here, and to prepare this Erratum.  All of the material has been reviewed and its correctness confirmed by the authors, who take full responsibility for the content.


\end{document}